\documentclass[conference]{IEEEtran}

\ifCLASSOPTIONcompsoc
  \usepackage[nocompress]{cite}
\else
  \usepackage{cite}
\fi

\usepackage{00_preamble}
\usepackage{00_macros}
\usepackage{00_spacing}

\begin{document}

\date{}

\title{\Large \bf SoK: How Sensor Attacks Disrupt Autonomous Vehicles: \\
    An End-to-end Analysis, Challenges, and Missed Threats}

\author{
{\rm Qingzhao Zhang$^{\dagger}$\textsuperscript{*}, Shaocheng Luo$^{\ddagger}$\textsuperscript{*}, Z. Morley Mao$^{\dagger}$, Miroslav Pajic$^{\ddagger}$, Michael K. Reiter$^{\ddagger}$}\\
$^{\dagger}$University of Michigan \qquad $^{\ddagger}$Duke University
}

\maketitle

\begin{abstract}

    Autonomous vehicles, including self-driving cars, ground robots, and drones, rely on multi-modal sensor pipelines for safe operation, yet remain vulnerable to adversarial sensor attacks. A critical gap is the lack of a systematic end-to-end view of how sensor-induced errors traverse interconnected modules to affect the physical world.
    To address this gap, we provide a comprehensive survey across platforms, sensing modalities, attack methods, and countermeasures. At its core is \Model (\modelAbbr), a graph-based framework that maps how attacks inject errors, the conditions for their propagation through modules from perception and localization to planning and control, and when they reach physical impact. From this analysis, our study distills 8 key findings that highlight the feasibility constraints of sensor attacks and uncovers 12 previously overlooked attack vectors exploiting inter-module interactions, several of which we validate via proof-of-concept experiments. We further conduct a pilot user study with autonomous vehicle security researchers, showing that \modelAbbr is viewed as a faithful and useful framework for qualitative system-level attack analysis. We also evaluate large language models (LLMs) for generating the analysis, demonstrating the potential of AI-powered automation.

\end{abstract}

\begingroup
\renewcommand\thefootnote{}\footnotetext{\textsuperscript{*}These authors contributed equally.}%
\endgroup

\section{Introduction} \label{sec:intro}

\begin{table*}[t]
    \centering
    \scriptsize
    \renewcommand{\arraystretch}{1.0} 
    \setlength{\tabcolsep}{4.0pt} 

    \caption{Comparison of our survey with existing works}
    \label{tab:survey_comparison}

    \begin{tabular}{|c|cccc|cccc|cccc|}
        \hline
        \multirow{2}{*}{\textbf{Ref}} 
        & \multicolumn{4}{c|}{\textbf{Attack Victim }} 
        & \multicolumn{4}{c|}{\textbf{Attack Location}} 
        & \multicolumn{4}{c|}{\textbf{Real-world Attack Feasibility Analysis}} 
       \\ 
        
        \cline{2-13}
        & \makecell{Specific \\ Sensors}
        & \makecell{Self-driving \\ Cars}  
        & \makecell{Ground \\ Vehicles}  
        & Drones  
        & \makecell{Perception} 
        & \makecell{Localization }         
        & \makecell{Planning}  
        & \makecell{Control}  
        & \makecell{System \\ Knowledge$^\dagger$}  
        & \makecell{Error \\ Propagation}  
        & \makecell{Effectiveness$^\ddagger$} 
        & \makecell{Generalizability$^\ast$}
        \\
        
        \hline
        {\cite{kim2024systematic}}  & \CIRCLE & \Circle & \LEFTcircle & \CIRCLE & \Circle  & \LEFTcircle  & \Circle  & \CIRCLE & \CIRCLE & \LEFTcircle  & \CIRCLE & \Circle   \\
        {\cite{shen2022sok}}  & \LEFTcircle & \CIRCLE & \Circle & \Circle & \CIRCLE  & \CIRCLE  & \CIRCLE  & \Circle  & \CIRCLE  & \Circle  & \CIRCLE  & \Circle \\
        {\cite{xu2023sok}}  & \CIRCLE & \CIRCLE & \CIRCLE & \CIRCLE & \LEFTcircle  & \LEFTcircle  & \LEFTcircle  & \LEFTcircle & \Circle & \Circle & \LEFTcircle  & \Circle  \\
        {\cite{kuhr2024sok}}  & \CIRCLE & \CIRCLE & \Circle & \Circle & \LEFTcircle  & \Circle  & \Circle  & \Circle & \Circle & \Circle & \CIRCLE  & \LEFTcircle  \\
        {\cite{altawy2016security}}  & \CIRCLE & \Circle & \Circle & \CIRCLE & \Circle  & \Circle  & \Circle  & \CIRCLE & \CIRCLE & \Circle & \LEFTcircle  & \Circle  \\
        {\cite{nassi2021sok}}  & \CIRCLE & \Circle & \Circle & \CIRCLE & \CIRCLE  & \Circle  & \Circle  & \CIRCLE & \CIRCLE & \Circle & \CIRCLE  & \Circle  \\
        {\cite{khan2026sok}}  & \CIRCLE & \CIRCLE & \Circle & \Circle & \CIRCLE  & \Circle  & \Circle  & \Circle & \CIRCLE & \LEFTcircle & \CIRCLE  & \Circle  \\
        \hline
        \textbf{Ours}  & \CIRCLE & \CIRCLE & \CIRCLE & \CIRCLE  & \CIRCLE  & \CIRCLE  & \CIRCLE  & \CIRCLE  & \CIRCLE  & \CIRCLE & \CIRCLE  & \CIRCLE \\
        \hline
    \end{tabular}
    
    \vspace{0.5em}
    {\scriptsize \CIRCLE\ = Elaborated; \LEFTcircle\ = Not explicitly stated or exhibits ambiguity; \Circle\ = Not stated;
    \, $^\dagger$ System knowledge - whether the attack model is white-box, grey-box, or black-box;\\
    $^\ddagger$ Effectiveness - whether an attack causes impactful system-level outcomes;
    \, $^\ast$ Generalizability - whether the attack is effective beyond environments in the original study.
    }
    
\end{table*}

Autonomous vehicles (AVs), such as self-driving cars (SDCs), unmanned ground vehicles (UGVs), and unmanned aerial vehicles (UAVs), rely on complex sensor pipelines to perceive their environments, estimate their states, and plan actions safely and efficiently in diverse conditions. They are widely deployed for urban mobility, industrial automation, and tasks like search and rescue. However, their reliance on multi-modal sensing, including localization sensors (e.g., Global Navigation Satellite System (GNSS), Inertial Measurement Units (IMUs)) and perception sensors (e.g., LiDARs, radars, cameras)~\cite{ayala2021sensors,van2018autonomous}, greatly expands the attack surface. Malicious interference such as GNSS spoofing, LiDAR injection, radar jamming, or electromagnetic attacks can mislead individual modules, causing perception failures, planning errors, and unsafe behaviors.

However, real safety risk arises from end-to-end effects: sensor perturbations matter only if they survive fusion and downstream modules to alter physical behavior. Component-level success does not guarantee system-level harm, so evaluating end-to-end feasibility and impact is essential. Yet prior surveys rarely provide this assessment (\tblref{tab:survey_comparison}). Many focus on a single modality such as cameras~\cite{kuhr2024sok}, emphasize the difficulty of the sensor attack itself~\cite{kim2024systematic}, or restrict scope to one vehicle type and miss cross-platform patterns~\cite{khan2026sok,nassi2021sok,altawy2016security,guo2017exploiting}. Xu et al.~\cite{xu2023sok} trace attack flows across system modules, but stay at module-level paths without assessing finer-grained characteristics of specific errors.

To bridge these gaps, our survey provides a comprehensive and structured analysis of sensor attack impacts across the full AV pipeline. First of all, we focus explicitly on the end-to-end modularized autonomous system pipeline, capturing the progression of error and uncertainty from sensing, through perception and localization, to planning and control. This pipeline-aware approach allows us to analyze how faults introduced at the sensor level can evolve and propagate to compromise downstream components and the physical world.
Building on this foundation, we broaden coverage across sensing modalities (camera, LiDAR, radar, IMU, GNSS, and fused systems) and platforms (SDC, UGV, UAV), enabling consistent multi-modal and cross-platform reasoning about feasibility and impact.
Together, the study addresses three research questions through the systematization of knowledge:

\begin{itemize}
    \item \textbf{RQ1: Propagation to impact.} How do sensor errors move through system modules (perception, localization, planning, etc.) to produce physical consequences?
    \item \textbf{RQ2: Feasibility barriers.} What common challenges impede end-to-end impactful attacks in realistic settings?
    \item \textbf{RQ3: Underexplored threats.} Which attack paths enabling end-to-end propagation remain overlooked, and how might they be realized?
\end{itemize}

Our systematization is organized using the proposed \Model (\modelAbbr), a graph-based framework that qualitatively organizes how sensor attack effects propagate through the AV pipeline. In \modelAbbr, the nodes represent errors at sensors or system components, the edges encode the conditions under which errors propagate, and the paths capture feasible routes from error injection to physical impact. Nodes also carry attributes (e.g., precision, continuity, environment dependence) that determine whether propagation continues downstream. This structure is used to visualize our answers to the three research questions: it makes propagation to impact explicit by enumerating conditioned paths (RQ1), reveals feasibility barriers by showing where attributes or transition conditions could challenge propagation (RQ2), and surfaces underexplored threats by highlighting unused or weakly constrained paths across modalities and platforms (RQ3). Although \modelAbbr is manually crafted rather than an automated tool, it distills insights from the literature and helps frame future research directions. We also evaluated LLMs' performance in \modelAbbr generation at the end of the study. In this work, we analyze end-to-end AV sensor attack feasibility across prior work, identifying key findings and overlooked attack vectors. Our main contributions are:
\begin{itemize}
    \item \textbf{Comprehensive survey of cross-platform and multi-modal sensor attacks.}
    We synthesize attacks on cameras, LiDARs, radars, IMUs, and GNSS across SDCs, UGVs, and UAVs, emphasizing implications for end-to-end attack feasibility.

    \item \textbf{Systematic analysis of error propagation and feasibility factors.}
    We introduce \modelAbbr to illustrate how sensor perturbations traverse perception, localization, tracking, prediction, planning, and control under explicit conditions (e.g., precision/continuity of errors), capturing qualitative factors that enable or block feasibility.

    \item \textbf{Identification of key insights and overlooked attacks.}
    Our analysis yields 8 insights on practical attack feasibility and 12 underexplored attack vectors exploiting inter-module interactions (several supported by proof-of-concept evidence).
\end{itemize}

\section{Preliminaries} \label{sec:preliminaries}

\begin{figure}[t]
    \centering
    \begin{subfigure}{0.99\linewidth}
    \centering
        \includegraphics[width=0.99\linewidth]{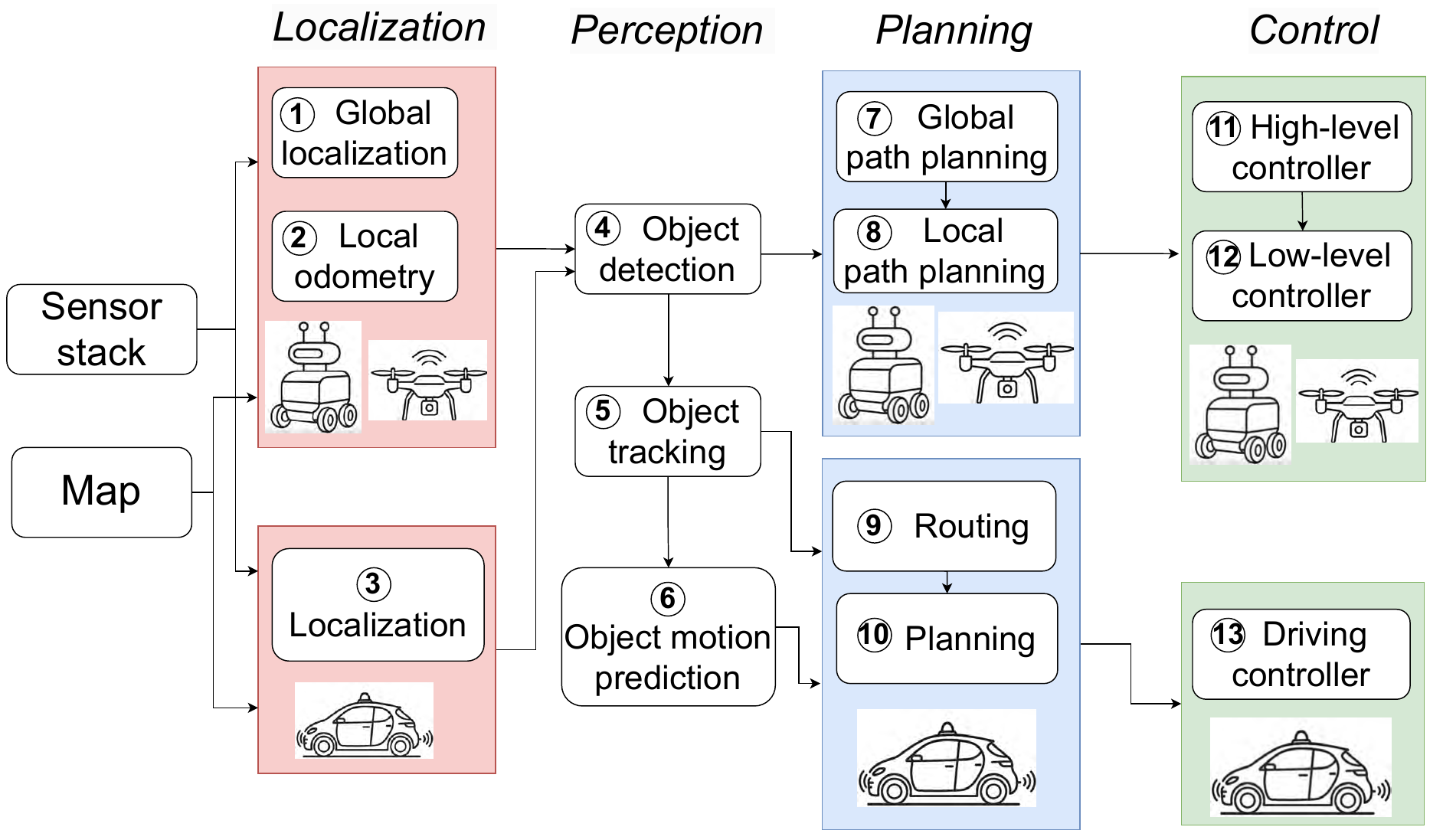}
    \end{subfigure}
    \caption{The system architecture of autonomous navigation in three representative platforms (SDCs, UGV, and UAVs).}
    \label{fig:system_diagram}
\end{figure}


As shown in \figref{fig:system_diagram}, SDCs, UGVs, and UAVs share a modular architecture of perception, localization, prediction, planning, and control. Heterogeneous sensors (e.g., cameras, LiDAR, radar, GPS, IMUs) supply environmental and ego-motion information to the localization and perception process for prediction and planning. Controllers then track the planned trajectories to navigate autonomously.

Despite this shared architecture, designs vary with platform constraints and navigation requirements. Robotic vehicles (RVs), i.e., ground and aerial platforms, typically operate in unstructured, self-organized environments and emphasize real-time localization, continuous trajectory planning, and local obstacle avoidance.
Global localization (\figref{fig:system_diagram}, \circled{1}) derives from GNSS or global occupancy maps \cite{schiotka2017robot}, while local odometry (\circled{2}) comes from wheel encoders and IMUs \cite{potokar2024robust}, often fused with visual \cite{scaramuzza2008appearance} and LiDAR odometry \cite{zhang2014loam}. 
RVs perceive the environment (\circled{4}) with local perceptual sensors, plan a path from the known goal and current location (\circled{7}), and adjust it dynamically to avoid unexpected obstacles (\circled{8}). 

In contrast, SDCs typically localize with GNSS or vision against HD maps (\circled{3}) \cite{jeong2020hdmi}, and their need for lane-level precision makes them rely heavily on multi-sensor fusion. Beyond the perception functionalities of RVs, SDCs add object tracking (\circled{5}) for car-following and motion prediction (\circled{6}) of nearby agents, to plan driving actions (\circled{10}) once a global route is determined (\circled{9}). 

These platforms also differ in control strategy and real-time requirements. UAVs depend heavily on IMU data for rapid stabilization, running low-level controllers (\circled{11}) at high frequency (e.g., 400 Hz in PX4 \cite{meier2015px4}) while high-level controllers (\circled{12}) guide trajectory following at 100-250 Hz. Ground vehicles execute the computed driving actions (\circled{10}) through lateral and longitudinal control, accounting for vehicle kinematics \cite{jo2015development}.

\section{Systematization Methodology}

\begin{table}[t]
    \scriptsize
    \centering
    \renewcommand{\arraystretch}{1.2} 
    \setlength{\tabcolsep}{3.0pt} 
    
    \caption{Node attribute schema of \modelAbbr.}
    \label{tbl:node_attributes}
    \begin{tabular}{|p{2cm}|p{6cm}|}
    \noalign{\global\arrayrulewidth1pt}\hline\noalign{\global\arrayrulewidth0.4pt}
    \thead{Attribute Name} & \thead{Description}  \\
    \noalign{\global\arrayrulewidth1pt}\hline\noalign{\global\arrayrulewidth0.4pt}
    \texttt{Precision} & \texttt{Low} – the error is uncontrolled; \texttt{Medium} – targeted error with considerable uncertainty; \texttt{High} – precise and reliable control over the injected error. \\
    \hline
    \texttt{Continuity} & \texttt{Low} – effective for only a single frame; \texttt{Medium} – impacts multiple frames but may not succeed consistently; \texttt{High} – reliably induces errors in every frame. \\
    \hline
    \texttt{Intensity} & \texttt{Low} – not observable or indistinguishable from benign faults; \texttt{Medium} – noticeable but with limited impact; \texttt{High} – significant deviations that exceed typical benign faults. \\
    \hline
    \texttt{DynamicTarget} & \texttt{True}/\texttt{False} - specifies whether the attack target (i.e., the victim vehicle) is moving or not. \\
    \hline
    \texttt{DynamicImpact} & \texttt{True}/\texttt{False} - indicates whether the injected error's value/state evolves over time, independent of how long the effect is sustained.\\
    \hline
    \texttt{Distance}- \texttt{ImpactToTarget} & The distance between the target (the victim vehicle) and the location of attack impact: \texttt{Low} – close enough to trigger an immediate response; \texttt{Medium} – nearby and will affect the target in the near future; \texttt{Far} – at a distance and does not directly influence the target. \\
    \noalign{\global\arrayrulewidth1pt}\hline\noalign{\global\arrayrulewidth0.4pt}
    \end{tabular}
\end{table}

\subsection{Scope of the Systematization}

This work focuses on sensor attacks on AV systems:
\begin{itemize}
    \item Sensor attacks that directly affect sensor measurements, and adversarial attacks that exploit sensor errors to disrupt system behavior, along with related countermeasures.
    \item Various autonomous platforms: SDCs,  UGVs, and UAVs.
    \item A range of sensors such as (stereo) cameras, LiDAR, mmWave radar, GNSS, and IMUs.
\end{itemize}

\noindent
The following aspects are out of scope for this study:
\begin{itemize}
    \item Unusual AV architectures such as foundation model stacks and end-to-end driving; we assume the canonical architecture in \figref{fig:system_diagram}.
    \item Network attacks on connected vehicles; we focus solely on single-vehicle sensor attacks.
    \item AI attacks not connected to autonomous vehicle systems.
    \item Stealthiness, which we discuss occasionally but do not address systematically.
\end{itemize}

\subsection{Literature Search}

We systematize the literature by searching Google Scholar and DBLP for work published between 2015–2025 in (1) major security venues: IEEE S\&P, CCS, USENIX Security, NDSS, AsiaCCS, EuroS\&P, DSN, RAID, ACSAC, and ESORICS; (2) system/software venues: MobiCom, MobiSys, SenSys, ICCPS, ICSE, ASE, and FSE; (3) AI/ML venues: ICCV, ECCV, CVPR, NeurIPS, and ICML; (4) robotics venues: ICRA, IROS, RSS, CoRL, and RA-L. Publications from other venues are included when relevant. We selected 121 papers for detailed systematization.


\subsection{\Model}
\label{sec:model}

We introduce the \Model\ (\modelAbbr) to present our analysis. Designed for systematization rather than formal verification, it uses a coarse schema and natural-language commentary to show how injected errors traverse system components and may culminate in safety-critical failures.


\myparagraph{Definitions}
\modelAbbr is a directed hypergraph $\graph = (\nodes, \transitions)$, where $\nodes$ is the set of nodes and $\transitions$ the set of transitions (hyperedges may have multiple source and target nodes). \tblref{tbl:collection_nodes} in the Appendix lists the nodes modeled by \modelAbbr.

Nodes in $\graph$ fall into three disjoint categories:
\begin{itemize}
    \item \emph{Physical Attack Vectors} ($\nodesAttack$): Entry points where an adversary injects perturbations or spoofing signals, such as image patches and spoofed sensor data.
    \item \emph{Errors of System Components} ($\nodesError$): Internal deviations from correct behavior that propagate across subsystems (e.g., planning) and affect downstream components.
    \item \emph{Physical-World Consequences} ($\nodesImpact$): Safety-critical violations manifesting in the real world, including collisions, off-road deviations, or erratic maneuvers.
\end{itemize}

Each node $\node \in \nodes$ is defined as a tuple:
\[
v = (\texttt{ID}, \texttt{Type}, \texttt{Attributes})
\]
where:
\begin{itemize}
    \item \texttt{ID}: A unique identifier for the node.
    \item \texttt{Type} $\in \{\texttt{attack}, \texttt{error}, \texttt{impact}\}$: The category of the node corresponding to the types above.
    \item \texttt{Attributes}: A fixed set of key–value fields describing the node’s properties; valid keys and value domains are listed in \tblref{tbl:node_attributes}. Supplementary natural-language descriptions are permitted when predefined attributes lack expressivity.
\end{itemize}

Transitions are categorized as:
\begin{itemize}
    \item \emph{Attack-to-Error}: How a physical attack vector induces a system error (e.g., LiDAR spoofing causing a false-positive obstacle).
    \item \emph{Error-to-Error}: How one internal error triggers another across functional modules (e.g., perception drift causing planning deviation).
    \item \emph{Error-to-Impact}: How an internal system error escalates to a physical-world violation (e.g., an invalid trajectory leading to a collision).
\end{itemize}

Each transition $\transition \in \transitions$ is a directed hyperedge connecting a set of \emph{source nodes} $\{\node_i\}$ to a set of \emph{target nodes} $\{\node_j\}$:
\[
t = (\texttt{ID}, \texttt{Condition})
\]
where:
\begin{itemize}
    \item \texttt{ID}: A unique identifier for the transition.
    \item \texttt{Condition}: A Boolean expression (or natural-language comment) defined over the attributes of the source nodes $\{\node_i\}$ and target nodes $\{\node_j\}$, specifying when propagation from source to target nodes is possible.
\end{itemize}

\myparagraph{Flexibility of using informal language} Because the attribute schema and Boolean predicates are intentionally coarse, we allow \emph{supplementary natural-language descriptions} for nodes or transition conditions when the formal attribute vocabulary lacks expressivity. These descriptions are for human interpretation and guidance, not machine execution. For instance, transitions involving assumptions of sensor fusion and physical-world scenarios are noted as ``camera dominates sensor fusion'' or ``depends on scenarios''.


\begin{table}[t]
    \scriptsize
    \centering
    \renewcommand{\arraystretch}{1.2} 
    \setlength{\tabcolsep}{1.0pt} 
    
    \caption{Selected rules of constructing \modelAbbr.}
    \label{tbl:graph_rules}
    \begin{tabular}{|c|p{2.0cm}|p{5.9cm}|}
    \noalign{\global\arrayrulewidth1pt}\hline\noalign{\global\arrayrulewidth0.4pt}
    \thead{ID} & \textbf{Rule Name} & \textbf{Description}  \\
    \noalign{\global\arrayrulewidth1pt}\hline\noalign{\global\arrayrulewidth0.4pt}
    R1 & Default inheritance & By default, the downstream error node will inherit attributes from upstream error nodes. \\
    \hline
    
    R2 & Uncertainty degrades precision & If the attack has uncontrollable real-world factors that harm accuracy, cap the node's attribute \texttt{Precision} to \texttt{Medium}. \\
    \hline
    
    R3 & Per-frame success indicates continuity & If the attack has a success rate under 90\% per frame, cap the node's attribute \texttt{Continuity} to \texttt{Medium}. \\
    \hline
    
    R4 & Tracking upgrades continuity & The object tracking by definition can upgrade node attribute \texttt{Continuity} from \texttt{Medium} to \texttt{High}. \\
    \hline

    R5 & Prediction extends positional error & Prediction extrapolates location errors into a longer-term future, potentially shrinking \texttt{DistanceTargetToImpact} from \texttt{Medium} to \texttt{Close}. \\
    \hline
    
    R6 & Tracking or prediction attacks require continuity, dynamic upstream errors & The transition to tracking error requires \texttt{Continuity} $=$ \texttt{Medium} while transition to prediction error requires \texttt{Continuity} $=$ \texttt{High}. Both transitions require \texttt{DynamicImpact}. \\
    \hline

    R7 & Sensor fusion as transition & Model fusion as a special transition that merges multiple sensor-error nodes into one detection/localization error node, conditioned on the attack error dominating the fusion so it carries downstream. \\
    \hline

    R8 & Scenario dependency in transitions & Transitions to planning error conditioned on object geometry or ego motion/intent are labeled as \emph{scenario-dependent}, i.e., feasible only in certain scenes. \\
    \hline

    R9 & Collision/brake requires a close-distance error & The transition to physical impact nodes like collision/brake requires \texttt{DistanceTargetToImpact} $=$ \texttt{Close} and \texttt{DynamicTarget}. \\
    
    \noalign{\global\arrayrulewidth1pt}\hline\noalign{\global\arrayrulewidth0.4pt}
    \end{tabular}
\end{table}

\myparagraph{Graph construction}
We construct \modelAbbr manually for each attack by reviewing prior work and applying AV domain expertise. Because many AV attacks share common architectures and error patterns, we distill general construction rules; selected rules appear in \tblref{tbl:graph_rules} and the full list in \tblref{tbl:collection_rules} in Appendix. They describe how attributes propagate (R1), how uncertainty and per-frame success cap \texttt{Precision} and \texttt{Continuity} (R2--R3), and how tracking and prediction can amplify or filter errors and change \texttt{Continuity} and \texttt{DistanceTargetToImpact} (R4--R6). We model sensor fusion explicitly as a transition (R7), capturing when an attack-induced error dominates or is suppressed by fusion, and encode scenario and proximity constraints for planning and physical impact nodes (R8--R9). These rules act as an extensible cookbook yielding consistent \modelAbbr instances across attacks, while still allowing expert judgment for undefined cases.

Manual construction can be error-prone and limited by individual expertise. To mitigate this, two domain experts independently construct \modelAbbr instances for each attack following the rules in \tblref{tbl:graph_rules}, then compare their graphs, resolve discrepancies through discussion, and reconcile them into a single agreed representation. Feedback from 15 additional domain experts in the pilot user study (\secref{sec:user_study}) is also integrated to correct \modelAbbrs. We separately tested LLMs' capability in generating such graphs (\secref{sec:llm}).

\myparagraph{One case study}
We illustrate \modelAbbr with an end-to-end LiDAR spoofing attack fabricating a lead vehicle and triggering unnecessary braking~\cite{cao2019adversarial,sun2020towards,hallyburton2022security,jin2023pla}. Feasibility has been demonstrated on certain hardware~\cite{sato2024lidar}, but practical constraints such as limited spoofable points, narrow injection angles, and next-generation LiDAR mitigations limit the impact. Spoofing close-range (\(<5\,\mathrm{m}\)) objects against modern sensors and robust detectors remains difficult.

\begin{figure}[t]
    \centering
    \includegraphics[width=0.99\linewidth]{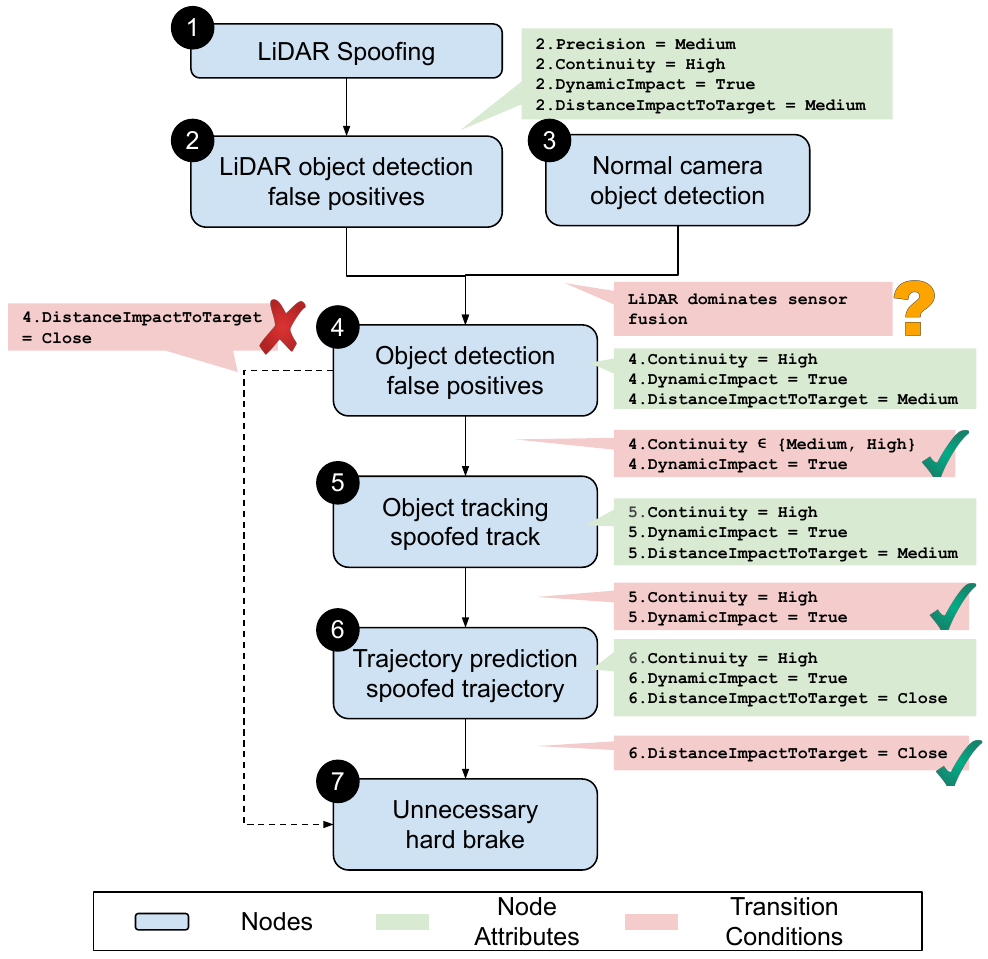}
    \caption{One example of \modelAbbr on LiDAR spoofing and object spoofing attacks in autonomous driving.}
    \label{fig:model_example}
\end{figure}

Given this background, we construct the \modelAbbr shown in \figref{fig:model_example} following the rules in \tblref{tbl:graph_rules}. The graph illustrates two attack paths originating from LiDAR spoofing.

The first path (\circlednum{1}\circlednum{2}\circlednum{4}\circlednum{7}) models a false positive from the LiDAR object detector, interpreted as a nearby ghost vehicle. However, triggering hard braking (\circlednum{7}) requires close-range impact (\texttt{DistanceImpactToTarget = Close}, R9 in \tblref{tbl:graph_rules}), technically difficult for current LiDAR spoofing. Thus, while conceptually correct, this path is not always practical due to hardware limitations.

The second path (\circlednum{1}\circlednum{2}\circlednum{4}\circlednum{5}\circlednum{6}\circlednum{7}) presents a more feasible alternative. Here, a mobile spoofed object creates a persistent spoofed track that propagates through tracking and trajectory prediction. Importantly, the prediction module can translate a medium-range spoofed track into a close-range predicted conflict (R5), causing the planner to execute hard braking. This path relies on continuous multi-frame spoofing and dynamic impact (R6), which are achievable with existing LiDAR spoofing techniques.

The model also captures sensor fusion, assuming the LiDAR-camera detection pipeline common in autonomous driving. Overall, it highlights both practical limitations of existing attacks and underexplored propagation paths that may enable realistic impacts.
\section{Literature Review and Systematization}
\label{sec:survey}

We systematize the literature on AV sensor attacks and countermeasures by analyzing their implications for error propagation across system modules, presented as \modelAbbrs, addressing \textbf{RQ1} (\secref{sec:intro}).

\subsection{Attacks}
\label{sec:survey_attacks}

\begin{table*}[ht]
\tiny
\centering
\caption{Summary of physical sensor attacks and their system-level implications.}
\label{tbl:attack_summary}
\renewcommand*{\arraystretch}{1.0}
\setlength{\tabcolsep}{1pt}
\begin{tabularx}{\textwidth}{|c|p{2.5cm}|p{1.5cm}|c|c|c|c|X|}
\noalign{\global\arrayrulewidth1pt}\hline\noalign{\global\arrayrulewidth0.4pt}
\thead{ID} & \thead{Attack Name} & \thead{References} & \thead{SYS} & \thead{Sensor} & \thead{Scene} & \thead{Access} & \thead{Summary of Error Propagation} \\
\noalign{\global\arrayrulewidth1pt}\hline\noalign{\global\arrayrulewidth0.4pt}

\attackid{camera_patch_object_spoofing} & Camera patch object spoofing & \cite{nassi2020phantom} & \VehicleIcon\UGVIcon & \CameraIcon & \IndoorIcon \OutdoorIcon & \Circle \LEFTcircle \CIRCLE & \AttackTag{Image patch} \To \ErrorTag{Camera detection false positives}\AttributeWarnTag{Continuity=Low} \TransitionWarnTag{Camera-only}\To \ErrorTag{Object detection false positives}\AttributeWarnTag{DistanceImpactToTarget=Close, DynamicTarget=True} \To \ErrorTag{Planning error} \To \ImpactTag{Braking} \\
\hline

\attackid{camera_patch_object_removal} & Camera patch object removal & \cite{wang2021dual,huang2020universal,shrestha2023towards} & \VehicleIcon\UGVIcon\DroneIcon & \CameraIcon & \IndoorIcon \OutdoorIcon & \Circle \LEFTcircle & \AttackTag{Image patch} \To \ErrorTag{Camera detection false negatives}\AttributeWarnTag{Continuity=High} \TransitionWarnTag{Camera-only}\To \ErrorTag{Object detection false negatives}\AttributeWarnTag{DistanceImpactToTarget=Close, DynamicTarget=True} \To \ErrorTag{Planning error} \To \ImpactTag{Collision} \\
\hline

\attackid{camera_sign_misdetection} & Camera sign misdetection & \cite{eykholt2018robust,guo2024invisible,kong2020physgan,xia2024moire,lovisotto2021slap} & \VehicleIcon & \CameraIcon & \OutdoorIcon & \Circle \LEFTcircle & \AttackTag{Image patch} \To \ErrorTag{Camera misdetection (sign)}\AttributeWarnTag{Continuity=High} \TransitionWarnTag{Depends on scenarios, no map data}\To \ErrorTag{Planning error} \To \ImpactTag{Rule violation} \\
\hline

\attackid{camera_patch_misclassification} & Camera patch misclassification & \cite{man2023person} & \VehicleIcon\UGVIcon & \CameraIcon & \OutdoorIcon & \Circle \LEFTcircle & \AttackTag{Image patch} \To \ErrorTag{Camera detection misclassification}\AttributeWarnTag{Continuity=High} \To \ErrorTag{Planning error} \To \ImpactTag{Unexpected driving decisions} \\
\hline

\attackid{camera_patch_tracking} & Camera patch tracking attack & \cite{jia2020fooling,muller2022physical,zhou2023f} & \VehicleIcon\UGVIcon & \CameraIcon & \OutdoorIcon & \Circle \LEFTcircle & \AttackTag{Image patch} \To \ErrorTag{Camera detection displacement}\AttributeWarnTag{Precision=High,Continuity=Medium} \TransitionWarnTag{Camera-only}\To \ErrorTag{Tracking hijacking}\AttributeWarnTag{Continuity=High} \To \ErrorTag{Prediction error} \To \ErrorTag{Planning error} \To \ImpactTag{Braking} OR \ImpactTag{Collision} \\
\hline

\attackid{camera_patch_slam} & Camera patch SLAM attack & \cite{davidson2016controlling,chen2024adversary,lou2025asymmetry} & \VehicleIcon\UGVIcon & \CameraIcon & \IndoorIcon \OutdoorIcon & \LEFTcircle & \AttackTag{Image patch}/\AttackTag{Strong light} \To \ErrorTag{Camera motion estimation error} \TransitionWarnTag{LiDAR dominates sensor fusion}\To \ErrorTag{Localization error}\AttributeWarnTag{Continuity=Medium/High} (\To \ImpactTag{Drifting}) OR (\TransitionWarnTag{No obstacle avoidance}\To \ImpactTag{Collision})  \\
\hline

\attackid{camera_patch_lane_detection} & Camera patch lane detection error & \cite{sato2021dirty,jing2021too} & \VehicleIcon\UGVIcon & \CameraIcon & \IndoorIcon \OutdoorIcon & \LEFTcircle & \AttackTag{Image patch} \To \ErrorTag{Lane detection error}\AttributeWarnTag{Continuity=High} \To \ErrorTag{Planning error} \To \ImpactTag{Off-road} \\
\hline

\attackid{camera_patch_depth} & Camera patch depth attack & \cite{cheng2022physical,xie2025flytrap,zheng2024physical,liu2024beware} & \VehicleIcon \DroneIcon & \CameraIcon & \OutdoorIcon & \Circle \LEFTcircle & \AttackTag{Image patch} \To \ErrorTag{Camera depth estimation error}\AttributeWarnTag{Precision=Medium, Continuity=Medium, Intensity=High} \To \ErrorTag{Planning error} \To \ImpactTag{Collision} OR \ImpactTag{Brake} \\
\hline

\attackid{camera_patch_latency} & Camera patch latency attack & \cite{muller2025investigating} & \VehicleIcon & \CameraIcon & \OutdoorIcon & \Circle \LEFTcircle & \AttackTag{Image patch} \TransitionWarnTag{Using projector; dark environment}\To \ErrorTag{Camera detection latency}\AttributeWarnTag{Intensity=High} \TransitionWarnTag{Camera dominates sensor fusion}\To \ErrorTag{Object detection false positives} \TransitionWarnTag{Depends on scenarios}\To \ErrorTag{Planning error} \To \ImpactTag{Collision} \\
\hline

\attackid{laser_camera_traffic_light} & Laser camera traffic light misdetection & \cite{yan2022rolling} & \VehicleIcon & \CameraIcon & \OutdoorIcon & \CIRCLE & \AttackTag{Laser} \To \ErrorTag{Camera image distortion} \To \ErrorTag{Camera traffic light misdetection}\AttributeWarnTag{Continuity=High} \TransitionWarnTag{Traffic light scenario}\To \ErrorTag{Planning error} \To \ImpactTag{Rule violation} \\
\hline

\attackid{ir_light_camera} & Infrared light camera attack & \cite{wang2021can,sato2024invisible} & \VehicleIcon & \CameraIcon & \OutdoorIcon & \CIRCLE & \AttackTag{Infrared light} \To \ErrorTag{Camera detection false positives}\AttributeWarnTag{Precision=Medium} OR \ErrorTag{Camera misdetection (traffic signal)} OR \ErrorTag{Localization deviation error} \TransitionWarnTag{Depends on scenarios}\To \ErrorTag{Planning error} \To \ImpactTag{Unexpected driving decisions} \\
\hline

\attackid{strong_light_camera_depth} & Strong light camera depth estimation attack & \cite{zhou2022doublestar} & \VehicleIcon \DroneIcon & \CameraIcon & \OutdoorIcon & \CIRCLE & \AttackTag{Strong light} \To \ErrorTag{Camera image distortion} \To \ErrorTag{Camera depth estimation error}\AttributeWarnTag{Precision=Medium,Continuity=Medium,Intensity=High} \To \ErrorTag{Planning error} \To \ImpactTag{Braking} OR \ImpactTag{Drifting} \\
\hline

\attackid{lidar_spoofing_object_spoofing} & LiDAR spoofing object spoofing & \cite{cao2019adversarial,sun2020towards,jin2023pla,hallyburton2022security,jin2025phantomlidar} & \VehicleIcon\UGVIcon & \LidarIcon (\CameraIcon) & \OutdoorIcon & \Circle \CIRCLE & \AttackTag{LiDAR spoofing} \To \ErrorTag{LiDAR detection false positives} \TransitionWarnTag{LiDAR dominates sensor fusion}\To \ErrorTag{Detection false positives}\AttributeWarnTag{Continuity=Medium, DistanceImpactToTarget=Close, DynamicTarget=True} \To \ErrorTag{Planning error} \To \ImpactTag{Braking} \\
\hline

\attackid{lidar_spoofing_object_removal} & LiDAR spoofing object removal & \cite{jin2023pla,cao2023you,jin2025phantomlidar} & \VehicleIcon\UGVIcon & \LidarIcon & \OutdoorIcon & \CIRCLE & \AttackTag{LiDAR spoofing} \To \ErrorTag{LiDAR detection false negatives} \TransitionWarnTag{LiDAR dominates sensor fusion}\To \ErrorTag{Object detection false negatives}\AttributeWarnTag{Continuity=High, DistanceImpactToTarget=Close, DynamicTarget=True} \To \ErrorTag{Planning error} \To \ImpactTag{Collision} \\
\hline

\attackid{lidar_spoofing_latency} & LiDAR spoofing latency attack & \cite{liu2023slowlidar} & \VehicleIcon & \LidarIcon & \OutdoorIcon & \Circle & \AttackTag{LiDAR spoofing} \To \ErrorTag{LiDAR detection latency}\AttributeWarnTag{Intensity=High} \TransitionWarnTag{LiDAR dominates sensor fusion}\To \ErrorTag{Object detection error} \TransitionWarnTag{Depends on scenario}\To \ErrorTag{Planning error} \To \ImpactTag{Collision} \\
\hline

\attackid{lidar_spoofing_slam} & LiDAR spoofing SLAM attack & \cite{fukunaga2024random,zhang2024prepared} & \VehicleIcon\UGVIcon & \LidarIcon & \OutdoorIcon & \CIRCLE & \AttackTag{LiDAR spoofing} \To \ErrorTag{LiDAR image distortion} \To \ErrorTag{LiDAR motion estimation error}\AttributeWarnTag{Continuity=Medium} \TransitionWarnTag{LiDAR dominates sensor fusion}\To \ErrorTag{Localization deviation error}\AttributeWarnTag{Continuity=Medium} \To \ImpactTag{Drifting} \\
\hline

\attackid{lidar_spoofing_odometry} & LiDAR spoofing odometry attack & \cite{song2025gradient} & \UGVIcon & \LidarIcon & \IndoorIcon \OutdoorIcon & \Circle & \AttackTag{LiDAR spoofing} \To \ErrorTag{LiDAR motion estimation error} \AttributeWarnTag{Continuity=Medium,Intensity=High} \To \ErrorTag{Planning error} \To \ImpactTag{Off-road} \\
\hline

\attackid{lidar_spoofing_localization} & LiDAR spoofing localization attack & \cite{nagata2025slamspoof,zhang2025prepared} & \UGVIcon & \LidarIcon(\CameraIcon) & \IndoorIcon & \Circle & \AttackTag{LiDAR spoofing} \To \ErrorTag{LiDAR motion estimation error} \To \ErrorTag{Localization deviation error} \AttributeWarnTag{Continuity=High,Intensity=Medium} \To \ErrorTag{Planning error} \To \ImpactTag{Off-road} \\
\hline

\attackid{radar_spoofing_object_spoofing} & mmWave radar spoofing object spoofing & \cite{sun2021control} & \VehicleIcon\UGVIcon & \RadarIcon & \IndoorIcon \OutdoorIcon & \CIRCLE & \AttackTag{Radar spoofing} \To \ErrorTag{Radar detection false positives} \TransitionWarnTag{Radar-only}\To \ErrorTag{Object detection false positives}\AttributeWarnTag{Continuity=Medium, DistanceImpactToTarget=Close, DynamicTarget=True} \To \ErrorTag{Planning error} \To \ImpactTag{Stalling} OR \ImpactTag{Braking} OR \ImpactTag{Lane change} \\
\hline

\attackid{radar_spoofing_object_removal} & mmWave radar spoofing object removal & \cite{hunt2024madradar, yan2016can} & \VehicleIcon\UGVIcon & \RadarIcon & \IndoorIcon \OutdoorIcon & \CIRCLE & \AttackTag{Radar spoofing} \To \ErrorTag{Radar detection false negatives} \TransitionWarnTag{Radar-only}\To \ErrorTag{Object detection false negatives}\AttributeWarnTag{Continuity=High, DistanceImpactToTarget=Close, DynamicTarget=True} \To  \ErrorTag{Planning error} \To \ImpactTag{Collision} \\
\hline

\attackid{gnss_spoofing_localization} & GNSS spoofing localization attack & \cite{shen2020drift,zhang2025ghost,zeng2018all} & \VehicleIcon & \GnssIcon & \OutdoorIcon & \CIRCLE & \AttackTag{GNSS spoofing} \To \ErrorTag{GNSS deviation error} \TransitionWarnTag{GNSS dominates the sensor fusion}\To \ErrorTag{Localization deviation error}\AttributeWarnTag{Continuity=High,Intensity=High} \To \ErrorTag{Routing Error} OR (\To \ErrorTag{Planning Error} \To \ImpactTag{Off-road}) \\
\hline

\attackid{gnss_spoofing_lidar_correction} & GNSS spoofing LiDAR correction attack & \cite{li2021fooling} & \VehicleIcon & \GnssIcon (\LidarIcon) & \OutdoorIcon & \Circle & \AttackTag{GNSS spoofing} \To \ErrorTag{LiDAR image distortion}\AttributeWarnTag{Precision=High} \To \ErrorTag{LiDAR detection error} \To \EllipsisTag  \\
\hline

\attackid{acoustic_emi_imu} & Acoustic/EMI IMU attack & \cite{jeong2023rocking,trippel2017walnut,jang2023paralyzing,son2015rocking} & \VehicleIcon\UGVIcon \DroneIcon & \ImuIcon & \IndoorIcon \OutdoorIcon & \CIRCLE & \AttackTag{Acoustic signal} \To \ErrorTag{IMU reading error}\AttributeWarnTag{Intensity=High} \TransitionWarnTag{IMU dominates sensor fusion}\To \ErrorTag{Planning error} \TransitionWarnTag{Depends on scenarios}\To \ImpactTag{Collision} \\
\hline

\attackid{acoustic_camera} & Acoustic signal camera attack & \cite{ji2021poltergeist,shamsi2024wip} & \VehicleIcon\UGVIcon \DroneIcon & \CameraIcon & \OutdoorIcon & \CIRCLE & \AttackTag{Acoustic signal} \To \ErrorTag{Camera image distortion}\AttributeWarnTag{Precision=Medium} \To \ErrorTag{Camera detection false positives} \TransitionWarnTag{Camera dominates sensor fusion}\To \ErrorTag{Object detection false positives} \To \EllipsisTag \\
\hline

\attackid{acoustic_patch_camera} & Acoustic + patch camera attack & \cite{zhu2023tpatch} & \VehicleIcon\UGVIcon & \CameraIcon & \OutdoorIcon & \Circle \LEFTcircle & (\AttackTag{Acoustic signal} \To \ErrorTag{Camera image distortion}\AttributeWarnTag{Precision=Medium}) AND \AttackTag{Image patch} \To \ErrorTag{Camera detection false positives}\AttributeWarnTag{Continuity=Medium} \TransitionWarnTag{Depends on scenarios, no map data}\To \ErrorTag{Planning error} \To \ImpactTag{Rule violation} \\
\hline

\attackid{physical_object_lidar_spoofing} & Physical object LiDAR object spoofing & \cite{yang2021robust,kobayashi2025invisible,zhu2021adversarial,zhang2022towards} & \VehicleIcon\UGVIcon & \LidarIcon & \OutdoorIcon & \Circle \CIRCLE & \AttackTag{Physical object} \To \ErrorTag{LiDAR detection false positives} \TransitionWarnTag{LiDAR dominates sensor fusion}\To \ErrorTag{Detection false positives}\AttributeWarnTag{Continuity=Medium, DistanceImpactToTarget=Close} \To \ErrorTag{Planning error} \To \ImpactTag{Braking} \\
\hline

\attackid{physical_object_lidar_removal} & Physical object LiDAR object removal & \cite{zhu2021can,tu2020physically,zhu2021adversarial,zhang2022towards} & \VehicleIcon\UGVIcon & \LidarIcon & \OutdoorIcon & \Circle \LEFTcircle & \AttackTag{Physical object} \To \ErrorTag{LiDAR detection false negatives} \TransitionWarnTag{LiDAR dominates sensor fusion}\To \ErrorTag{Object detection false negatives}\AttributeWarnTag{Continuity=High, DistanceImpactToTarget=Close} \To \ErrorTag{Planning error} \To \ImpactTag{Collision} \\
\hline

\attackid{physical_patch_radar_measurement} & Physical patch mmWave radar range/angle/speed attack & \cite{chen2023metawave} & \VehicleIcon\UGVIcon & \RadarIcon & \OutdoorIcon & \LEFTcircle & \AttackTag{Physical patch} \To \ErrorTag{Radar range measurement error} \TransitionWarnTag{Radar dominates sensor fusion}\To \ErrorTag{Object detection false positives}\AttributeWarnTag{Continuity=High, DistanceImpactToTarget=Close} \TransitionWarnTag{Depends on scenario}\To \ErrorTag{Planning error} \To \ImpactTag{Braking} OR \ImpactTag{Collision} \\
\hline

\attackid{physical_patch_radar_removal} & Physical patch mmWave radar object removal & \cite{zhu2023tilemask} & \VehicleIcon\UGVIcon & \RadarIcon & \OutdoorIcon & \Circle \LEFTcircle & \AttackTag{Physical patch} \To \ErrorTag{Radar detection false negatives} \To \ErrorTag{Object detection false negatives}\AttributeWarnTag{Continuity=High, DistanceImpactToTarget=Close} \To  \ErrorTag{Planning error} \To \ImpactTag{Collision} \\
\hline

\attackid{physical_object_disappearing} & Physical object disappearing (camera \& LiDAR) & \cite{cao2021invisible,zhu2024malicious,abdelfattah2021adversarial} & \VehicleIcon\UGVIcon & \CameraIcon \LidarIcon & \OutdoorIcon & \Circle & \AttackTag{Physical object} \To \ErrorTag{Camera detection false negatives} AND \ErrorTag{LiDAR detection false negatives} \TransitionWarnTag{LiDAR/camera sensor fusion}\To \ErrorTag{Object detection false negatives}\AttributeWarnTag{Continuity=High, DistanceImpactToTarget=Close} \To \ErrorTag{Planning error} \To \ImpactTag{Collision} \\
\hline

\attackid{physical_object_depth} & Physical object depth estimation attack & \cite{zheng2024pi} & \VehicleIcon\UGVIcon & \CameraIcon & \OutdoorIcon & \CIRCLE & \AttackTag{Physical object} \To \ErrorTag{Depth estimation error} \TransitionWarnTag{Depth-based obstacle avoidance}\To \ErrorTag{Object detection displacement}\AttributeWarnTag{Continuity=Medium, DistanceImpactToTarget=Close} \To \ErrorTag{Planning error} \To \ImpactTag{Collision} OR \ImpactTag{Braking} \\
\hline

\attackid{physical_vehicle_prediction} & Physical vehicle prediction attack & \cite{zhang2022adversarial,cao2022advdo,song2023acero} & \VehicleIcon\UGVIcon & NA & \OutdoorIcon & \Circle \LEFTcircle \CIRCLE & \AttackTag{Physical object} \To \ErrorTag{Prediction displacement error}\AttributeWarnTag{Precision=Medium, Continuity=Medium} \TransitionWarnTag{Depends on scenarios}\To \ErrorTag{Planning error} \To \ImpactTag{Braking} OR \ImpactTag{Collision} \\
\hline

\attackid{physical_object_perception_prediction} & Physical object perception-prediction attack & \cite{lou2024first,yu2025enduring} & \VehicleIcon\UGVIcon & \LidarIcon & \OutdoorIcon & \Circle \LEFTcircle & \AttackTag{Physical object} \To \ErrorTag{LiDAR detection false positives} \TransitionWarnTag{LiDAR dominates sensor fusion}\To \ErrorTag{Object detection displacement}\AttributeWarnTag{Precision=Medium, Continuity=Medium} \To \ErrorTag{Prediction displacement error}\AttributeWarnTag{Precision=Medium, Continuity=Medium} \TransitionWarnTag{Depends on scenarios}\To \ErrorTag{Planning error} \To \ImpactTag{Braking} OR \ImpactTag{Acceleration} OR \ImpactTag{Lane change} \\
\hline

\attackid{physical_object_planning} & Physical object planning attack & \cite{wan2022too} & \VehicleIcon\UGVIcon & NA & \OutdoorIcon & \Circle \LEFTcircle & \AttackTag{Physical object} \TransitionWarnTag{Depends on scenarios}\To \ErrorTag{Planning error} \To \ImpactTag{Braking} \\

\noalign{\global\arrayrulewidth1pt}\hline\noalign{\global\arrayrulewidth0.4pt}

\end{tabularx}

{\scriptsize 
\textbf{SYS}: \VehicleIcon\,SDCs, \UGVIcon\,UGVs, \DroneIcon\,UAVs. 
\textbf{Scene}: \IndoorIcon\, indoor, \OutdoorIcon\, outdoor.
\textbf{Feasibility Threats}: Attributes affecting feasibility; red denotes fatal ones invalidating the \modelAbbr path. \\
\textbf{Sensor}: includes the main target sensor and optional additional sensor modalities considered (in parentheses); 
\CameraIcon\, camera, \LidarIcon\, LiDAR, \RadarIcon\, mmWave radar, \GnssIcon\, GNSS, \ImuIcon\, IMU. \\
\textbf{Access}: 
\Circle\, white-box (known model parameters and system details), 
\LEFTcircle\, gray-box (known data/system configurations, or transfer attacks), 
\CIRCLE\, black-box (query access only). \\
\textbf{Graph presentation}:
\ErrorTag{Node} is node in \modelAbbr.  
\AttributeWarnTag{A=B} and \TransitionWarnTag{X} are node attribute and transition condition that are required by the attack and is challenging to achieve.
}

\end{table*}

\tblref{tbl:attack_summary} summarizes the analyzed attacks: target platforms, affected sensors, scenarios, attacker knowledge, and representative \modelAbbrs. It highlights key node attributes and transition conditions influencing attack feasibility; detailed \modelAbbrs are in \figref{fig:all_sepg_graphs} in the Appendix. We generate one \modelAbbr per paper but present one representative graph per category for clarity. Building on this, we categorize attack vectors into direct physical attacks on sensor hardware (\secref{sec:results_sensor_attack}) and adversarial attacks manipulating sensor data (\secref{sec:results_adversarial_attacks}).

\subsubsection{Immediate Effects of Sensor Attacks}
\label{sec:results_sensor_attack}

Physical sensor attacks exploit hardware vulnerabilities to inject errors into the autonomy stack. They directly corrupt sensor readings and can serve as a prerequisite for adversarial attacks.

\myparagraph{Camera-based attacks (\attacksref{camera_patch_object_spoofing,strong_light_camera_depth})} In autonomous driving, physical adversarial patches (e.g., stickers) on traffic signs or vehicles can make objects misclassified or undetected by camera-based detectors, leading to rule violations or safety risks~\cite{eykholt2018robust,zhao2019seeing,jia2020fooling,man2023person}. Projectors can cast adversarial images onto vehicles, roads, or structures for similar detection errors~\cite{muller2022physical,muller2025investigating}; SLAP further turns the projection on/off in short bursts to bypass anomaly detectors~\cite{lovisotto2021slap}.
In vision-based localization and SLAM, adversarial patches can corrupt visual cues for pose estimation~\cite{chen2024adversary}. In drones, downward-facing cameras used for optical flow~\cite{beauchemin1995computation} can be perturbed by projected light or lasers, creating false stationary features and drift, especially at low altitudes or indoors~\cite{davidson2016controlling}. Infrared light, strong light, or lasers can affect camera images~\cite{wang2021can,zhou2022doublestar,yan2022rolling,sato2024invisible}.

Adversarial camera image patches could induce misclassifications~\cite{man2023person}, object detection false positives or false negatives~\cite{eykholt2018robust,zhao2019seeing}, displacement errors of detected objects~\cite{jia2020fooling,muller2022physical}, increased latency of camera-based perception~\cite{muller2025investigating}, and localization errors~\cite{chen2024adversary}. Precision and continuity of such attacks are generally challenging when the victim vehicle and the target object are moving (attributes \texttt{DynamicTarget} and \texttt{DynamicImpact}) and far apart (attribute \texttt{DistanceImpactToTarget}), because mobility and distance change the camera view at runtime while the patch is generated offline and static. Prior work~\cite{zhao2019seeing} proved the attack is hard but still possible in such scenarios, by leveraging expectation over transformation (EoT) when optimizing adversarial examples.
Camera-based attacks also depend on scenarios, inherently affected by lighting.

\myparagraph{LiDAR spoofing (\attacksref{lidar_spoofing_object_spoofing,lidar_spoofing_localization})}
LiDAR spoofing injects fake points into a victim’s scan by emitting carefully timed laser pulses that mimic legitimate returns. External laser sources, often modulated and steered via rotating mirrors or galvanometers, are tuned to match the wavelength, pulse width, and timing of commercial LiDARs~\cite{sato2024lidar}. Jin et al.~\cite{jin2025phantomlidar} further leverage intentional electromagnetic interference (IEMI) to target multiple LiDAR components, enabling both point cloud manipulation and power-off. Attackers can thus create ghost vehicles or obstacles~\cite{cao2019adversarial,sun2020towards,hallyburton2022security} or remove objects by overwriting or masking legitimate returns~\cite{cao2023you,jin2023pla}. Because LiDAR also serves localization and SLAM, spoofing can disrupt these functions~\cite{fukunaga2024random}.

LiDAR spoofing can cause detection false positives and false negatives~\cite{cao2019adversarial,sun2020towards,hallyburton2022security,cao2021invisible,jin2023pla}, increase perception latency~\cite{liu2023slowlidar}, and induce localization errors~\cite{fukunaga2024random}. Its effectiveness, however, is constrained by view angle and distance (especially for close, forward objects), point sparsity and limited spoofing resolution, and emerging hardware defenses such as anomaly rejection, timing checks, and cryptographic encoding~\cite{sato2024lidar}. The precision of injected errors (\texttt{Precision}) and the effective range relative to the victim LiDAR (\texttt{DistanceImpactToTarget}) are thus restricted, and adverse weather (fog, heavy rain, snow) further degrades both LiDAR performance and attacks.

\myparagraph{mmWave radar spoofing (\attacksref{radar_spoofing_object_spoofing,radar_spoofing_object_removal})}
mmWave radar spoofing injects crafted waveforms mimicking the victim radar to induce false positives (FPs)~\cite{hunt2024madradar}, false negatives (FNs) via jamming~\cite{yan2016can,hunt2024madradar}, translation of detected objects~\cite{hunt2024madradar}, and angle-of-arrival (AoA) errors~\cite{sun2021control}; passive meta-material tags can also create FPs and FNs~\cite{chen2023metawave}. Because mmWave radars underpin blind-spot detection (BSD), automatic emergency braking (AEBS), lane-change assist (LCA), and rear traffic alert (RTA)~\cite{waldschmidt2021automotive}, spoofing directly threatens core safety features. It can be white-box~\cite{sun2021control,miura2019low,komissarov2021spoofing,chauhan2014platform} or black-box~\cite{hunt2024madradar,vennam2023mmspoof}, depending on whether the attacker knows waveform parameters; black-box attacks approximate chirp period, slope, and frame duration from observed frames to predict future transmissions and sustain the attack.

mmWave radar spoofing induces object detection and localization errors~\cite{hunt2024radcloud}, but its effectiveness depends on several preconditions. FP/FN attacks are constrained by the distance between spoofer and victim due to signal attenuation \linebreak (\texttt{DistanceImpactToTarget}), and FN attacks require saturating the victim radar with strong spoofing signals~\cite{hunt2024madradar,yan2016can}. AoA attacks demand precise control of spoofing angles (\texttt{Precision}), and attacks such as translation require continuously injecting a sequence of spoofed signals (\texttt{Continuity}).

\myparagraph{GNSS spoofing (\attacksref{gnss_spoofing_localization,gnss_spoofing_lidar_correction})}
GNSS spoofing is a physical-layer attack in which an adversary transmits forged satellite signals to manipulate a receiver’s perceived position or time. Spoofed signals are typically generated via SDRs to closely mimic authentic GNSS transmissions, and must be aligned with legitimate signals in frequency, phase, and timing to succeed~\cite{humphreys2008assessing}. As GNSS (e.g., GPS) is a key localization input, such attacks can mislead autonomous systems, causing path deviations or unsafe decisions. ``Carry-off'' attacks gradually overpower genuine signals to provide seamless but false navigation data~\cite{humphreys2008assessing}, and have been shown to disrupt localization in autonomous and robotic vehicles, leading to off-road deviations or driving rule violations~\cite{shen2020drift}. GNSS outputs are also used to support LiDAR calibration, particularly for compensating motion-induced distortions, enabling GNSS spoofing to indirectly manipulate LiDAR-based detection~\cite{li2021fooling}.

GNSS spoofing targets localization primarily, but real-world effectiveness is constrained: precision is often limited to meter-level by environmental noise and partial knowledge of target hardware, and modern autonomy stacks fuse GNSS with IMUs, LiDAR odometry, and visual SLAM, reducing impact. It therefore requires prerequisites and remains limited in precision~\cite{kim2024systematic}.

\myparagraph{Acoustic/EMI signals on MEMS-based sensors (\attacksref{acoustic_emi_imu,acoustic_patch_camera})}
Acoustic and electromagnetic interference (EMI) attacks exploit physical vulnerabilities in MEMS-based sensors such as accelerometers, gyroscopes, and even vision systems. High-frequency or resonant acoustic waves from directional speakers or ultrasonic transducers couple into mechanical structures and produce spurious vibrations misinterpreted as motion, yielding false acceleration or angular-rate readings~\cite{trippel2017walnut}. Shamsi et al.~\cite{shamsi2024wip} further use high-power pulsed lasers to excite acoustic vibrations in MEMS gyroscopes, though this remains lab-only. EMI works similarly: injected signals couple into sensor circuitry, power rails, or signal lines, corrupting inertial measurements or communication channels~\cite{jang2023paralyzing}. Such signals can corrupt the motion sensing of vehicles or drones~\cite{trippel2017walnut,son2015rocking} and even manipulate control actuators in CPS~\cite{tu2018injected}. Acoustic interference with inertial sensors can also disrupt camera stabilization, causing blurred images and detection failures~\cite{ji2021poltergeist,zhu2023tpatch}. Together, they degrade perception, localization, and control in autonomy stacks.

While these attacks can induce both perception and localization errors, their real-world feasibility is limited. They require precise alignment and close proximity between source and target sensor, and are highly sensitive to environmental factors such as background noise, material damping, sensor mounting, shielding, and grounding, all of which undermine reliability. Sensor-fusion mechanisms further diminish their impact on deployed systems.

\myparagraph{Physical objects or agent behavior (\attacksref{physical_object_lidar_spoofing,physical_object_planning})}
Beyond direct sensor spoofing, adversaries can passively mislead autonomous systems by manipulating the environment or agent motion rather than tampering with sensors, inducing perception and prediction errors. For example, 3D-printed adversarial meshes can evade LiDAR or camera detectors~\cite{tu2020physically,cao2021invisible,lou2024first}, reflective structures can fool LiDAR models~\cite{zhu2021can}, commodity objects (cardboard, signs) can mislead LiDAR semantic segmentation~\cite{zhu2021adversarial}, and improved adversarial shapes offer greater robustness~\cite{zhu2024ae}. Training-time data poisoning can further inject a backdoor into the LiDAR detector~\cite{zhang2022towards}. Dynamic driving patterns can generate adversarial trajectories that cause large prediction errors and unsafe decisions~\cite{zhang2022adversarial,cao2022advdo,song2023acero}, while adversarial objects at strategic locations can degrade depth estimation of occluded objects~\cite{zheng2024pi}.

These attacks correspond to nodes for physical objects or agent behaviors that induce object detection and trajectory prediction errors. A key challenge is precise, reproducible physical execution: patterns or trajectories optimized offline under idealized conditions mismatch at deployment. Their effectiveness is thus limited by \texttt{Precision} (how closely real behavior matches the optimized pattern) and \texttt{Continuity} (how long the induced error is sustained).

\subsubsection{Adversarial Attacks Amplifying Errors}
\label{sec:results_adversarial_attacks}

Adversarial attacks exploit learning-based components in the autonomy stack to amplify physical sensor errors into downstream failures, enabling end-to-end attacks with real-world impact. In \modelAbbr, they correspond to transitions propagating errors through the system stack.

\myparagraph{Object detection attacks (\attacksref{camera_patch_object_spoofing,camera_patch_misclassification}, \attacksref{camera_patch_lane_detection,lidar_spoofing_latency}, \attacksref{radar_spoofing_object_spoofing,radar_spoofing_object_removal}, \attackref{gnss_spoofing_lidar_correction}, \attacksref{acoustic_camera,physical_object_depth})}
Adversarial attacks on object detection manipulate perception outputs by adding minimal, carefully crafted perturbations to sensor data, exploiting deep learning model vulnerabilities. They often overcome inherent sensor limitations: adversarial point cloud attacks can spoof objects with only a few synthetic points, bypassing LiDAR spoofing’s challenge of injecting large point volumes across wide views~\cite{sun2020towards,cao2019adversarial}; adversarial patches use Expectation-over-Transformation (EoT) to stay effective under varying distances and perspectives, mitigating camera-based attacks’ sensitivity to environmental changes~\cite{zhao2019seeing}; the passive mmWave radar attack optimizes the placement of physical patches in limited sizes to fool DNN-based detection models~\cite{zhu2023tilemask}. These methods enable reliable, stealthy deception in physical sensor attacks.

To succeed against more advanced perception systems, these attacks must also contend with defenses like sensor fusion, which integrates multiple sensors to improve robustness and accuracy and reduce the impact of single-sensor perturbations. To counter this, Cao et al.~\cite{cao2021invisible} proposed a physical object attack that simultaneously perturbs camera and LiDAR so the vehicle misdetects the object. Hallyburton et al.~\cite{hallyburton2022security} manipulate only LiDAR but make it dominate the fused result by carefully designing the geometric region to attack. These strategies increase the attack success.

Acoustic interference with camera stabilizers is another vector against object detection (\secref{sec:results_sensor_attack}). Zhu et al.\cite{zhu2023tpatch} introduced a conditional attack combining acoustic signals with adversarial patches: detection is unaffected by the patch alone but errs when both are applied, enabling a stealthy, triggerable attack.
Additionally, Li et al.~\cite{li2021fooling} use GPS spoofing to introduce subtle errors into LiDAR calibration, corrupting the resulting point clouds and degrading LiDAR-based object detection.

\myparagraph{Object tracking attacks (\attackref{camera_patch_tracking})}
Adversarial attacks on object tracking cause identity switches, track fragmentation, or false tracks, exploiting vulnerabilities in tracking algorithms, particularly those relying on deep learning and data association.
Jia et al.~\cite{jia2020fooling} introduced tracker hijacking: adversarial examples causing consistent detection displacement over a few frames hijack the tracker onto incorrect trajectories.
Müller et al.\cite{muller2022physical} achieved a similar effect physically, projecting adversarial patterns into the environment to mislead Siamese-based trackers.
Wang et al.~\cite{wang2024physical} manipulated a physical object’s trajectory to induce identity switches in multi-object tracking, revealing a novel vulnerability.

These attacks are modeled as transitions from detection errors (e.g., displacement) to tracking errors (e.g., identity switches or false tracks), all represented as nodes in $\nodesError$. Most assume camera-based detection, leaving their effectiveness on multi-modal fusion systems uncertain, and they require sustained perturbations to detections of moving objects (attributes \texttt{Continuity} and \texttt{DynamicImpact}). Whether tracking errors reach physical impacts such as collisions or hard braking depends on driving context; e.g., the perturbed track must be close to and interacting with the victim vehicle (attribute \texttt{DistanceImpactToTarget}).

\myparagraph{Motion/trajectory prediction attacks (\attackref{physical_vehicle_prediction}, \attackref{physical_object_perception_prediction})}
Adversarial attacks on motion and trajectory prediction mislead forecasts of surrounding agents’ future movements, directly affecting planning and decision-making.
Zhang et al.\cite{zhang2022adversarial} showed an adversary can drive a real vehicle along a crafted trajectory that makes the victim predict incorrectly and decide unsafely.
Cao et al.\cite{cao2022advdo} improved the naturalness and reproducibility of such trajectories, evading detection while still misleading prediction.
Song et al.\cite{song2023acero} developed a black-box approach modeling ego--agent interactions to generate adversarial maneuvers effective across prediction algorithms.
Lou et al.\cite{lou2024first} gave the first physical-world demonstration, placing objects around a parked car to induce LiDAR detection errors that caused prediction failures and unsafe behaviors such as unnecessary hard braking to yield to a stationary vehicle.

They are represented as adversarial behaviors propagating \linebreak perception-induced errors into prediction errors. For attacks using real vehicles to execute adversarial trajectories~\cite{zhang2022adversarial,cao2022advdo,song2023acero}, the key challenge is the \texttt{Precision} of physically reproducing optimized trajectories in the real world. Moreover, existing works demonstrate success in isolated frames but lack analysis of \texttt{Continuity}, raising concerns about sustaining the adversarial effect over time. To induce impacts such as collisions or hard braking, the erroneous predictions must interact directly with the victim vehicle, requiring a relatively low \texttt{DistanceImpactToTarget}.

\myparagraph{Localization attacks (\attackref{camera_patch_slam}, \attackref{lidar_spoofing_slam}, \attackref{gnss_spoofing_localization}, \attackref{acoustic_emi_imu})}
Adversarial attacks on localization can significantly amplify initial sensor errors by exploiting weaknesses in multi-sensor fusion or visual SLAM. Shen et al.~\cite{shen2020drift} induce localization drift by opportunistically spoofing GPS at moments when its input can override more reliable modalities, achieving off-road deviations under fusion-based localization. For visual SLAM, Chen et al.~\cite{chen2024adversary} use near-invisible adversarial patches captured by the camera, causing substantial mapping and pose estimation errors. Lou et al.~\cite{lou2025asymmetry} extend this to online HD map construction, showing that a single roadside flashlight or adversarial patch can corrupt the constructed map.

Such attacks follow different propagation paths (camera patches or GPS spoofing) and assume different architectures (GPS/IMU/LiDAR localization or vision SLAM), inheriting the feasibility challenges of the corresponding sensor attacks; e.g., attacking fusion by corrupting only GPS is confronted with \texttt{Precision} challenge.

\myparagraph{Planning attacks (\attackref{physical_object_planning})}
Such attacks target decision-making modules, manipulating high-level path selection or low-level motion plans to induce unsafe or inefficient behaviors. For example, Wan et al.~\cite{wan2022too} systematically discover planning denial-of-service vulnerabilities, where objects at selected locations make the planner generate overly conservative maneuvers such as braking and detouring.
They propagate from upstream perception or localization errors to planning, and their feasibility depends heavily on scenarios due to reliance on specific maps, rules, or planner heuristics.

\begin{table*}[ht]
\tiny
\centering
\caption{Summary of sensor attack defense mechanisms and their implications on systematic error propagation.}
\label{tbl:defense_summary}
\renewcommand*{\arraystretch}{1.0}
\setlength{\tabcolsep}{1pt}
\begin{tabularx}{\textwidth}{|c|p{2cm}|p{3cm}|c|c|c|X|p{3.5cm}|}
\noalign{\global\arrayrulewidth1pt}\hline\noalign{\global\arrayrulewidth0.4pt}
\thead{ID} & \thead{Defense Name} & \thead{References} & \thead{SYS} & \thead{Sensor} & \thead{Scene} & \thead{Implication on Error Propagation} & \thead{Targeted attacks} \\
\noalign{\global\arrayrulewidth1pt}\hline\noalign{\global\arrayrulewidth0.4pt}

\defenseid{hardware_raw_signal} & Hardware or raw signal level defenses & GNSS~\cite{liu2021stars,jansen2016multi,oligeri2020gnss,oligeri2019drive,sathaye2022semperfi}, LiDAR~\cite{sato2024lidar}, radar~\cite{sun2021control,moon2022bluefmcw}, IMU~\cite{trippel2017walnut,jeong2023rocking} & \makecell[t]{\VehicleIcon\UGVIcon\\\DroneIcon} & \makecell[t]{\LidarIcon\GnssIcon\\\RadarIcon\ImuIcon} & \IndoorIcon\OutdoorIcon & Directly limiting the \texttt{Intensity}, \texttt{Precision}, and \texttt{Continuity} of the physical attack node, e.g., \AttackTag{LiDAR spoofing}, \AttackTag{GNSS spoofing}, \AttackTag{Radar spoofing} and \AttackTag{Acoustic signal (IMU)}. & LiDAR attacks \attacksref{lidar_spoofing_object_spoofing,radar_spoofing_object_spoofing}, GNSS attacks \attacksref{radar_spoofing_object_spoofing,radar_spoofing_object_removal}, radar attacks \attacksref{gnss_spoofing_localization,gnss_spoofing_lidar_correction}, IMU attack \attackref{acoustic_emi_imu}  \\
\hline

\defenseid{perception_fusion} & Perception multi-sensor fusion & \cite{zhu2023understanding,cheng2024fusion,wang2024mmcert,huang2025commit,liang2022bevfusion,li2022deepfusion,le2024diffusion,yue2025roburcdet,liu2021seeing} & \VehicleIcon\UGVIcon & \makecell[t]{\CameraIcon\LidarIcon\\\RadarIcon} & \IndoorIcon\OutdoorIcon & Enforcing a condition on the transition from separate sensor object detection errors to the fused object detection errors. The condition limits \texttt{Intensity}, \texttt{Precision}, and \texttt{Continuity} of the fused object detection error. & 3D object detection \& depth estimation attacks \attackref{camera_patch_object_spoofing}, \attackref{camera_patch_object_removal}, \attackref{camera_patch_misclassification}, \attackref{camera_patch_tracking}, \attackref{camera_patch_depth}, \attackref{ir_light_camera}, \attackref{strong_light_camera_depth}, \attackref{lidar_spoofing_object_spoofing}, \attackref{lidar_spoofing_object_removal}, \attackref{gnss_spoofing_localization}, \attackref{gnss_spoofing_lidar_correction}, \attacksref{acoustic_camera,physical_object_depth} \\
\hline

\defenseid{localization_fusion} & Localization multi-sensor fusion & \cite{broumandan2018spoofing,shan2020lio,he2025ligo,gao2024fast,wu2023lidar,shen2023lateral} & \makecell[t]{\VehicleIcon\UGVIcon\\\DroneIcon} & \LidarIcon\GnssIcon\ImuIcon & \IndoorIcon\OutdoorIcon & Enforcing a condition on the transition from separate sensor odometry/localization errors to the fused localization errors. The condition limits \texttt{Intensity}, \texttt{Precision}, and \texttt{Continuity} of the fused localization error. & Localization/SLAM attacks \attackref{camera_patch_slam}, \attackref{lidar_spoofing_slam}, \attackref{radar_spoofing_object_spoofing}, \attackref{acoustic_emi_imu} \\
\hline

\defenseid{ai_robustness} & AI model robustness improvements & camera~\cite{cheng2024self,liu2022segment,pathak2024model,kang2024diffender}, LiDAR~\cite{sun2020towards}, prediction~\cite{zhang2022adversarial,cao2023robust,pourkeshavarz2024cadet,jiao2023semi,pourkeshavarz2024dyset} & \makecell[t]{\VehicleIcon\UGVIcon\\\DroneIcon} & \CameraIcon\LidarIcon & \IndoorIcon\OutdoorIcon & Reducing \texttt{Intensity}, \texttt{Precision}, and \texttt{Continuity} of the error in AI model outputs, such as \ErrorTag{object detection error}, \ErrorTag{Prediction error}, \ErrorTag{Lane detection error}, etc. & AI adversarial attacks \attacksref{camera_patch_object_spoofing,camera_patch_depth}, \attacksref{laser_camera_traffic_light,lidar_spoofing_object_removal}, \attackref{radar_spoofing_object_removal}, \attacksref{acoustic_camera,physical_object_lidar_removal}, \attacksref{physical_object_disappearing,physical_object_perception_prediction} \\
\hline

\defenseid{consistency_checks} & Perception spatial/temporal/contextual consistency checks & camera~\cite{han2024visionguard,nassi2020phantom,zhang2022all,lin2024phade}, LiDAR~\cite{cho2023adopt,sun2020towards,hau2021shadow,kobayashi2025invisible,xiao2023exorcising}, multi-modal~\cite{xu2024physcout,muller2024vogues} & \VehicleIcon\UGVIcon & \CameraIcon\LidarIcon & \OutdoorIcon & Anomaly detection and mitigation leveraging limited \texttt{Precision} and \texttt{Continuity} of induced perception errors including the \ErrorTag{object detection error} and \ErrorTag{object tracking error}. & Object detection and tracking attacks \attacksref{camera_patch_object_spoofing,camera_patch_tracking}, \attacksref{lidar_spoofing_object_spoofing,lidar_spoofing_object_removal}, \attacksref{radar_spoofing_object_removal,gnss_spoofing_lidar_correction}, \attacksref{acoustic_camera,physical_object_disappearing}, \attackref{physical_object_perception_prediction} \\
\hline

\defenseid{control_software} & Control software level defenses & \cite{dash2024specguard,kim2022drivefuzz,choi2020cyber,choi2018detecting,quinonez2020savior,dash2024diagnosis} & \makecell[t]{\VehicleIcon\UGVIcon\\\DroneIcon} & NA & \IndoorIcon\OutdoorIcon & Directly limiting \texttt{Intensity}, \texttt{Precision}, and \texttt{Continuity} of the \ErrorTag{Planning error}. & Generally applicable to \attacksref{camera_patch_object_spoofing,physical_object_planning} but could only defend against significant malicious behaviors. \\
\hline

\defenseid{multi_agent} & Multi-agent collaboration & GPS~\cite{jansen2018crowd}, camera~\cite{hallyburton2025trust}, LiDAR~\cite{wang2025threat,zhang2024data}, sensor-agnostic~\cite{yeke2025automated,hallyburton2025security} & \makecell[t]{\VehicleIcon\UGVIcon\\\DroneIcon} & \makecell[t]{\CameraIcon\LidarIcon\\\GnssIcon} & \OutdoorIcon & Depends on the collaboration scheme, suppressing the \texttt{Intensity}, \texttt{Precision}, and \texttt{Continuity} of the corresponding error node. & Generally applicable to \attacksref{camera_patch_object_spoofing,physical_object_planning} but relies on multi-agent communication schemes. \\

\noalign{\global\arrayrulewidth1pt}\hline\noalign{\global\arrayrulewidth0.4pt}

\end{tabularx}

\end{table*}

\subsection{Defenses}
\label{sec:survey_defenses}

We systematize defenses by their impact on error propagation, i.e., which inter-module transitions they block or mitigate. Because each category spans a broad research area, we summarize its mainstream approach and refer readers to surveys for further detail. \tblref{tbl:defense_summary} summarizes our systematization.

\myparagraph{Hardware or raw data level defenses (\defenseref{hardware_raw_signal})}
These defenses harden front ends or vet raw signals. For GNSS, independent physical references, multi receiver consistency, and hardened or inertially coupled receivers raise the bar for coherent spoofing~\cite{liu2021stars,jansen2016multi,oligeri2020gnss,oligeri2019drive,sathaye2022semperfi,broumandan2018spoofing}. LiDAR deploys next generation timing randomization and pulse fingerprinting~\cite{sato2024lidar}; radar uses randomized frequency trajectories~\cite{moon2022bluefmcw}; IMU protections mitigate resonant, acoustic, or electromagnetic signals via isolation and model based denoising~\cite{trippel2017walnut,jeong2023rocking}. In \modelAbbr, they act on the earliest attack nodes, constraining intensity, precision, and continuity of injected signals before they propagate, yet remain modality specific and can be undermined by adaptive, stronger spoofers, e.g., attacks on next generation LiDAR~\cite{hayakawa2025breaking,sato2024lidar}. See prior surveys for detail~\cite{wu2020spoofing,salguero2024state,boukabou2024cybersecurity}.

\myparagraph{Multi-sensor fusion (\defensesref{perception_fusion,localization_fusion})}
Perception and localization routinely fuse heterogeneous sensors to improve robustness. Perception stacks integrate camera, LiDAR, and mmWave radar, and deep learning–based fusion uses learnable architectures over multi-modal data for better performance and robustness~\cite{liang2022bevfusion,li2022deepfusion}; AI certification has been applied to such systems for provable guarantees against attacks~\cite{huang2025commit,wang2024mmcert}. Localization commonly combines IMU, GNSS, and LiDAR \cite{he2025ligo,gao2024fast,wu2023lidar}. IMUs stabilizing drones are vulnerable to acoustic resonance~\cite{son2015rocking,trippel2017walnut} and electromagnetic interference (EMI)~\cite{jang2023paralyzing}, so flight controllers like Pixhawk integrate multiple IMUs with different resonance characteristics; recent work also fuses IMUs with tachometers~\cite{meng2025mars} or high-frequency indoor positioning~\cite{tu2019flight} to compensate compromised sensors.
Fusion increases the system robustness against single-sensor attacks, proved by benchmarking results~\cite{zhu2023understanding,jin2024unity}, yet adaptive strategies can still bias joint estimates and corrupt fused outputs \cite{cheng2024fusion,hallyburton2022security,cao2021invisible}. Understanding how fusion reshapes attack surfaces and error propagation is therefore essential. Readers are referred to comprehensive prior surveys for details~\cite{wang2019multi,elsanhoury2021survey}.

\myparagraph{AI model improvements (\defenseref{ai_robustness})}
Defenses strengthen model robustness via redesigned architectures, robust training, and test-time enhancements. Models encoding temporal or spatial relations improve stability under distribution shifts and can defend against sensor attacks~\cite{sun2020towards,qin2023unifusion}; adversarial training and data augmentation apply across tasks including detection and prediction \cite{cheng2024self,zhang2022adversarial,cao2023robust}; test-time processing can suppress perturbations with smoothing or clipping \cite{zhang2022adversarial,wang2024mmcert,liu2022segment}. In \modelAbbr, these measures reduce the intensity, precision, and continuity of output errors at the corresponding error nodes, though many techniques are uncertified and adaptive attacks remain significant challenges. See prior survey~\cite{girdhar2023cybersecurity,ibrahum2024deep}.

\myparagraph{Spatial, temporal, or contextual consistency checks (\defenseref{consistency_checks})}
Because many attacks lack either \texttt{Precision} or \texttt{Continuity}, their outputs violate short-horizon spatial, temporal, or semantic relations. Consistency defenses use this gap as anomaly signals: cameras flag patch attacks via temporal incoherence or scene context mismatches~\cite{han2024visionguard,nassi2020phantom}; LiDAR checks laser physics, occlusion, and temporal stability~\cite{cho2023adopt,sun2020towards,hau2021shadow,kobayashi2025invisible,xiao2023exorcising}; and sensor agnostic methods scrutinize trajectory smoothness and physically plausible motion~\cite{xu2024physcout,muller2024vogues}. In practice, these checks most effectively bound perception errors in outdoor settings where rich environmental features supply many cross constraints.

\myparagraph{Control software level defenses (\defenseref{control_software})}
These methods detect, diagnose, and repair unsafe planner or controller behaviors at runtime, using specification guided monitoring, fuzzing guided test generation, program analysis, and recovery mechanisms \cite{dash2024specguard,kim2022drivefuzz,choi2020cyber,choi2018detecting,quinonez2020savior,dash2024diagnosis}. In \modelAbbr, they act on the planning error node by directly limiting the intensity, precision, and continuity of erroneous control outputs before they reach actuators. They are broadly applicable across platforms, though they are most effective against significant or persistent deviations rather than subtle manipulations.

\myparagraph{Multi-agent collaboration as defenses (\defenseref{multi_agent})}
Collaborative sche-mes use cross-agent agreement to identify and correct malicious or corrupted data, for example, crowd-based GPS validation \cite{jansen2018crowd}, trust-aware camera sharing \cite{hallyburton2025trust}, LiDAR consistency and threat assessment across peers \cite{wang2025threat,zhang2024data}, and sensor-agnostic consensus frameworks \cite{yeke2025automated,hallyburton2025security}. In \modelAbbr, such collaboration suppresses the affected GNSS sensing and object detection error nodes by exploiting spatial overlap and viewpoint diversity, but effectiveness depends on communication quality and trust assumptions.

\section{Findings and Insights}
\label{sec:findings}

Based on the systematization, we present findings on the feasibility of AV sensor attacks and propose underexplored threats, addressing \textbf{RQ2} and \textbf{RQ3} respectively.

\subsection{Analysis of End-to-End Attack Feasibility}
\label{sec:results_findings}

\begin{finding}[label=finding:sensor_fusion]
    Multimodal sensor fusion poses a significant challenge to the effectiveness of single-sensor attacks; however, it is not sufficiently analyzed in the existing adversarial attack literature.
\end{finding}

As discussed in \secref{sec:survey_defenses} (\defenseref{perception_fusion} in \tblref{tbl:defense_summary}), multi-modal fusion is widely used to mitigate single-sensor uncertainty. 
When a single-sensor attack ignores fusion, the added redundancy challenges the \texttt{Intensity}, \texttt{Precision}, and \texttt{Continuity} of the resulting error: an object removed from LiDAR data may still be detected by the camera, nullifying the attack.
Sensor fusion is not a certified method, and the arm race between adaptive attacks and defenses underscores the need for deeper security analysis~\cite{cheng2024fusion,khan2026sok}.
Notably, only 4 of the 59 object-detection or localization attacks in \tblref{tbl:attack_summary} were evaluated under sensor fusion, making systematic and easy-to-use fusion-aware attack assessment an important future direction.


\begin{finding}[label=finding:precision_challenge]
    Attack precision often degrades due to environmental situations and the attacker’s limited knowledge. Ensuring robustness under such uncertainty is challenging but essential for effective attack design.
\end{finding}

Precision is how accurately an attacker can induce a specific error in the victim system. It is essential for subtle, real-time manipulations but hard to achieve in practice: weather, lighting, sensor noise, and occlusion distort how perturbations are sensed, and attackers rarely know the victim's exact sensor setup, algorithms, or state, which adds uncertainty.

Robust attack design seeks to close the gap between intended and actual errors. Patch-based image attacks optimize over angle, distance, illumination, and motion~\cite{zhao2019seeing,eykholt2018robust,sato2021dirty}; physical objects for LiDAR are optimized to reduce the gap between expected and captured points~\cite{zhu2024ae}; and when an attack is prepared offline, designers optimize over possible motion paths for robustness~\cite{lou2024first,chen2024adversary}.

Despite these efforts, two precision challenges remain. 
(1) Physical sensor attacks are inherently uncertain, so downstream modules should not assume the manipulated data exactly matches the target state. LiDAR attacks~\cite{cao2019adversarial,liu2023slowlidar} produce optimized point clouds in simulation, but real world spoofing reliability is unclear; laser, strong light, and infrared attacks suffer alignment and aiming issues~\cite{wang2021can,yan2022rolling,zhou2022doublestar}; and trajectory prediction attacks~\cite{zhang2022adversarial,cao2022advdo} require precise control to reproduce the planned path (\secref{sec:results_sensor_attack}). 
(2) Limited attacker knowledge hinders fine-grained attacks. Many studies assume white-box access (\tblref{tbl:attack_summary}), while black-box variants are usually weaker, and some assumptions are unrealistic for a remote adversary. For example, FLAT~\cite{li2021fooling} (\attackref{gnss_spoofing_lidar_correction}) optimizes GPS spoofing using the victim LiDAR image, unlikely to be available to the attacker, casting doubt on feasibility.

Conversely, defenses can exploit imperfect precision (and continuity) for anomaly detection (\defenseref{consistency_checks} in \tblref{tbl:defense_summary}), provided the induced uncertainty is rigorously modeled and separable from benign variability; formalizing it may enable generalizable defenses.

\begin{finding}[label=finding:continuity_challenge]
    Certain attacks require high continuity to be effective, as their impact relies on being sustained across consecutive frames.
\end{finding}

Continuity is an attacker’s ability to inject errors across consecutive frames. It is critical because AVs often employ resilient components, such as Kalman filter–based sensor fusion and tracking, to correct transient faults, whereas sustained errors force repeated incorrect decisions, significantly amplifying an attack’s impact.

Certain attacks explicitly require continuity. Object tracking attacks~\cite{jia2020fooling,muller2022physical} rely on persistent detection errors to gradually shift an object’s estimated position, and trajectory prediction attacks~\cite{zhang2022adversarial,cao2022advdo,song2023acero,lou2024first} similarly depend on sustained displacement errors in detected or tracked objects. While prior work shows continuous detection errors are feasible, their reliability is sensitive to environmental noise and dynamic conditions. Object removal attacks~\cite{wang2021dual,huang2020universal,jin2023pla,cao2023you} also require continuity, as tracking modules tolerate occasional false negatives. GPS spoofing~\cite{shen2020drift} often starts with subtle location errors that grow over time to bypass Kalman filter–based fusion while staying stealthy, making continuous spoofing the main bottleneck.

Continuity is also essential for specific physical impacts. Object spoofing against a camera or LiDAR must persist over multiple frames to be perceived as a static obstacle triggering unnecessary braking or evasive maneuvers, and tracking and prediction attacks fabricate behaviors such as a fake lane change that force inappropriate braking or swerving; in both cases the fabrication must stay consistent long enough to affect planning. Zhang et al.~\cite{zhang2022adversarial} showed this difficulty: sustaining prediction errors over six frames is substantially harder than the single-frame attack.

\begin{finding}[label=finding:scenario_challenge]
    The proximity of injected object detection errors to the victim vehicle critically influences physical effects like hard braking or collisions.
\end{finding}


To trigger dangerous behaviors like hard braking, evasive maneuvers, or collisions, a misleading object must appear close enough to the vehicle’s path to force immediate response. Previous studies do not necessarily guarantee such physical-world feasibility, especially when the distance to the target is not carefully considered.

In fact, injecting an error close to the victim vehicle poses additional challenges. LiDAR spoofing is harder at close range because dense point clouds require precise signal matching, and feasibility depends heavily on the LiDAR hardware and spoofing equipment used~\cite{sato2024lidar}. Physical-object LiDAR attacks face similar problems, as close-range objects produce more detailed point clouds that are harder to fake. While Zhu et al.~\cite{zhu2021can} (\attackref{physical_object_lidar_removal}) demonstrated object removal as close as 5 meters on certain LiDAR setups, effectiveness dropped as the object moved closer. Similarly, laser, strong light, or infrared attacks on cameras~\cite{yan2022rolling,wang2021can,zhou2022doublestar} (\attacksref{laser_camera_traffic_light,strong_light_camera_depth}) can distort images but struggle at close range, since nearby objects occupy more pixels and richer visual features that resist distortion.

\begin{finding}[label=finding:simplified_system]
    End-to-end system implementation goes beyond simply connecting AI models. Using an oversimplified system for evaluation may not accurately reflect real-world attack effectiveness.
\end{finding}

Researchers commonly connect object detection, tracking, and prediction algorithms to approximate an AV perception stack. While often representative, a gap remains between such pipelines and real systems, whose specific designs can significantly influence how errors propagate during attacks, as shown below.

First, real system stacks select among several algorithms according to the current scenario, a design adopted by both industry-level open-source stacks, Baidu Apollo~\cite{apollo} and Autoware~\cite{autoware}. Apollo's trajectory prediction, for example, uses the neural predictor proven vulnerable by previous studies~\cite{zhang2022adversarial} (\attackref{physical_vehicle_prediction}) only when the object is close and the road has no clear lane marks; otherwise, lane information dominates the prediction. This challenges the generality of trajectory prediction attacks.

Second, even simple tolerance mechanisms significantly affect attack outcomes. Because object detection inherently includes noise and errors, Apollo smooths trajectories before prediction and may disable prediction entirely if an object is not moving fast enough, mitigating the instability of prediction algorithms. The physical object prediction attack~\cite{lou2024first} (\attackref{physical_object_perception_prediction}) assumes minor perception errors for a parked car propagate into large prediction errors; yet in Apollo, such a static object would not be predicted. These designs deserve careful consideration in security analyses.

Such software level design can also serve as a defense (\defenseref{consistency_checks} in \tblref{tbl:defense_summary}). Existing work mines and enforces invariants in control software but rarely couples this with security analysis of potential attacks; more intelligent and generalizable security-focused software auditing is promising.

\begin{finding}[label=finding:scenario_dependency]
Errors that indirectly alter planning decisions (e.g., traffic sign misdetection, latency of detection, prediction errors) cause the attack impacts (e.g., braking or collisions) only in carefully designed scenarios.
\end{finding}

This finding concerns how upstream errors translate into planning errors. Most direct perception/localization faults have obvious effects on planning: a spoofed object ahead forces a stop, a removed nearby obstacle can induce a collision, and localization drift can steer the vehicle off-road. These are relatively straightforward to exploit, whereas \emph{indirect} errors require more carefully crafted conditions. We capture this scenario dependence as the transition condition to planning errors.

For instance, traffic-sign misdetection (\attackref{camera_sign_misdetection}, \attackref{acoustic_patch_camera}) is usually demonstrated with stop signs, whose safety consequence is intuitive, while misdetecting other signs is less explored. Detection latency (\attackref{camera_patch_latency}, \attackref{lidar_spoofing_latency}) depends on relative speeds between colliding vehicles, and prediction attacks hinge on specific situations such as fake lane changes/merges or wrong acceleration estimates. The literature commonly notes highly scenario-dependent success rates (e.g., errors on closer targets more directly affect planning)~\cite{zhang2022adversarial,song2023acero}.

These cases highlight the need to evaluate attack generality and feasibility at the scenario level, i.e., how easily an attacker can create or opportunistically encounter vulnerable scenes. Conversely, real-time scenario risk assessment strengthens defenses by allocating more defensive resources to high-risk situations.

\subsection{Under-explored Potential Attacks}
\label{sec:new_attacks}

\begin{table*}[t]
\tiny
\centering
\caption{Summary of potential new attacks.}
\label{tbl:new_attacks}
\renewcommand*{\arraystretch}{1.1}
\setlength{\tabcolsep}{2pt}
\begin{tabularx}{\textwidth}{|c|p{3.5cm}|c|c|c|c|X|}
\noalign{\global\arrayrulewidth1pt}\hline\noalign{\global\arrayrulewidth0.4pt}
\thead{ID} & \thead{Attack Name} & \thead{SYS} & \thead{Sensor} & \thead{Scene} & \thead{Access} & \thead{Summary of Error Propagation} \\
\noalign{\global\arrayrulewidth1pt}\hline\noalign{\global\arrayrulewidth0.4pt}

\newattackid{tracking_by_misclassification} & Tracking attack by misclassfication (camera or LiDAR) & \VehicleIcon\UGVIcon\DroneIcon & \CameraIcon \LidarIcon & \OutdoorIcon & \Circle \LEFTcircle & (\AttackTag{Image patch} \To \ErrorTag{Camera misclassification}) AND/OR (\AttackTag{LiDAR spoofing}  \To \ErrorTag{LiDAR misclassification}) \TransitionWarnTag{LiDAR/camera sensor fusion}\To \ErrorTag{Object detection misclassification}\AttributeWarnTag{Continuity=Medium} \To \ErrorTag{Tracking broken tracks}\AttributeWarnTag{Continuity=Medium} \To \ErrorTag{Prediction error} \TransitionWarnTag{Depends on scenarios}\To \ErrorTag{Planning error} \To \ErrorTag{Collision} \\
\hline

\newattackid{detection_latency_discrepancy} & Detection latency discrepancy attack (camera and LiDAR) & \VehicleIcon\UGVIcon & \CameraIcon \LidarIcon & \OutdoorIcon & \Circle \LEFTcircle & (\AttackTag{Image patch} \To \ErrorTag{Camera detection latency}\AttributeWarnTag{Intensity=Medium}) AND/OR (\AttackTag{LiDAR spoofing} \To \ErrorTag{LiDAR detection latency}\AttributeWarnTag{Intensity=Medium}) \TransitionWarnTag{LiDAR/camera sensor fusion}\To \ErrorTag{Object detection error} \To ... \\
\hline

\newattackid{prediction_by_localization} & Prediction attack by localization error & \VehicleIcon & \GnssIcon \ImuIcon & \OutdoorIcon & \Circle \LEFTcircle \CIRCLE & (\AttackTag{GNSS spoofing} \To \ErrorTag{GNSS error}) AND/OR (\AttackTag{Acoustic signal} \To \ErrorTag{IMU error}) \TransitionWarnTag{IMU/GNSS sensor fusion}\To \ErrorTag{Localization error}\AttributeWarnTag{Precision=Medium, Continuity=Medium, Intensity=Medium} \To \ErrorTag{Coordination transformation error} \To \ErrorTag{Trajectory prediction error} \TransitionWarnTag{Depends on scenarios}\To \ErrorTag{Planning error} \To \ImpactTag{Collision} OR \ImpactTag{Braking} \\
\hline

\newattackid{lidar_spoofing_tracking} & LiDAR spoofing tracking attack & \VehicleIcon\UGVIcon & \LidarIcon & \OutdoorIcon & \Circle \LEFTcircle \CIRCLE & \AttackTag{LiDAR spoofing} \To \ErrorTag{LiDAR detection displacement} \TransitionWarnTag{LiDAR dominates sensor fusion}\To \ErrorTag{Object detection displacement}\AttributeWarnTag{Precision=Medium, Continuity=Medium} \To \ErrorTag{Tracking error} \To \ErrorTag{Prediction error}\AttributeWarnTag{DistanceImpactToTarget=Close} \TransitionWarnTag{Depends on scenarios}\To \ErrorTag{Planning error} \To \ImpactTag{Brake} OR \ImpactTag{Collision} \\
\hline

\newattackid{lidar_spoofing_misclassification} & LiDAR spoofing misclassification & \VehicleIcon\UGVIcon & \LidarIcon & \IndoorIcon \OutdoorIcon & \Circle \LEFTcircle & \AttackTag{LiDAR spoofing} \To \ErrorTag{LiDAR detection misclassification} \TransitionWarnTag{LiDAR dominates sensor fusion}\To \ErrorTag{Object detection misclassification}\AttributeWarnTag{Continuity=High} \TransitionWarnTag{Depends on scenarios}\To \ErrorTag{Planning error (wrong object type)} \To \ImpactTag{Improper driving behavior} \\
\hline

\newattackid{physical_object_misclassification} & Physical object misclassification (camera \& LiDAR) & \VehicleIcon\UGVIcon & \CameraIcon \LidarIcon & \OutdoorIcon & \Circle \LEFTcircle & \AttackTag{Physical object} \To \ErrorTag{Camera misclassification} AND \ErrorTag{LiDAR misclassification} \TransitionWarnTag{LiDAR/camera sensor fusion}\To \ErrorTag{Object detection misclassification}\AttributeWarnTag{Continuity=High} \TransitionWarnTag{Depends on scenarios}\To \ErrorTag{Planning error (wrong object type)} \To \ImpactTag{Collision} \\
\hline

\newattackid{joint_detection_tracking_prediction} & Joint detection-tracking-prediction attack (camera or LiDAR) & \VehicleIcon & \CameraIcon \LidarIcon & \OutdoorIcon & \Circle \LEFTcircle & \AttackTag{Image patch} AND/OR \AttackTag{LiDAR spoofing} \TransitionWarnTag{Sensor fusion}\To \ErrorTag{Object detection error}\AttributeWarnTag{Precision=Medium, Continuity=Medium} \To \ErrorTag{Tracking error} \To \ErrorTag{Prediction error} \TransitionWarnTag{Depends on scenarios}\To \ErrorTag{Planning error} \To \ImpactTag{Collision} OR \ImpactTag{Braking} \\
\hline

\newattackid{acoustic_camera_tracking} & Acoustic/electromagnetic signal camera tracking attack & \VehicleIcon\UGVIcon\DroneIcon & \CameraIcon & \OutdoorIcon & \CIRCLE & \AttackTag{Acoustic/electromagnetic signal} \To \ErrorTag{Camera blur} \To \ErrorTag{Camera detection error} \TransitionWarnTag{Camera dominates sensor fusion}\To \ErrorTag{Object detection error}\AttributeWarnTag{Precision=Medium, Precision=Medium} \To \ErrorTag{Tracking lost track} OR \ErrorTag{Tracking ID switch} \To \ImpactTag{Denial of service} \\
\hline

\newattackid{joint_camera_lidar_detection} & Joint camera + LiDAR object detection attacks & \VehicleIcon\UGVIcon & \CameraIcon \LidarIcon & \IndoorIcon \OutdoorIcon & \Circle \LEFTcircle \CIRCLE & A cluster of attacks coordinate camera attacks and LiDAR attacks to combat sensor fusion. \\
\hline

\newattackid{joint_gnss_imu_localization} & Joint GNSS + IMU localization attacks & \VehicleIcon\UGVIcon\DroneIcon & \GnssIcon \ImuIcon & \OutdoorIcon & \CIRCLE & A cluster of attacks that coordinate GNSS spoofing attacks and acoustic IMU attacks to combat sensor fusion. \\
\hline

\newattackid{joint_perception_planning} & Joint perception and planning attacks & \VehicleIcon\UGVIcon & \CameraIcon \LidarIcon \RadarIcon & \OutdoorIcon & \Circle \LEFTcircle \CIRCLE & A cluster of attacks that inject perception errors smartly to exploit planning vulnerabilities (e.g., placing off-road ghost objects to stop vehicles, combining object spoofing and planning attack A34). \\
\hline

\newattackid{scenario_aware_hybrid} & Scenario-aware longer-term hybrid attack & \VehicleIcon\UGVIcon\DroneIcon & Any & \IndoorIcon\OutdoorIcon & \Circle \LEFTcircle \CIRCLE & A cluster of attacks that leverages a sequence of different sensor attacks to achieve a targeted impact (e.g., continuous object spoofing and prediction attack blocking a vehicle at the intersection permanently). \\

\noalign{\global\arrayrulewidth1pt}\hline\noalign{\global\arrayrulewidth0.4pt}
\end{tabularx}


\end{table*}


Building on the error‑propagation analysis above, we identify under‑explored adversarial attacks in \tblref{tbl:new_attacks}, detailing a few as below.

\myparagraph{Novel combinations of sensor modality and error type} Attacks are not limited to a single sensor or failure mode; mixing modalities and compounding errors creates new threat paths. As summarized in \tblref{tbl:new_attacks}, camera misclassification and tracking attacks extend to LiDAR (\newattacksref{lidar_spoofing_tracking,physical_object_misclassification}); coordinated multi-sensor attacks can undermine sensor fusion (\newattacksref{joint_camera_lidar_detection,joint_gnss_imu_localization}); and planning attacks via physical objects can be replicated with sensor-level object spoofing (\newattackref{joint_perception_planning}). Attackers can also chain attacks to elicit targeted AV behaviors in specific scenarios (\newattackref{scenario_aware_hybrid}). We highlight Attacks \newattackref{tracking_by_misclassification} and \newattackref{detection_latency_discrepancy}.

\begin{attack}[label=attack:misclassification_to_tracking]{\newattackref{tracking_by_misclassification}}
Tracking attack by misclassification. By launching a perception misclassification attack, such as using adversarial image patches, the attacker causes an object to be repeatedly identified as different classes over time. This inconsistency disrupts temporal coherence, degrading tracking and prediction performance.
\end{attack}

The first example links camera/LiDAR misclassification to tracking errors, to which tracking is highly sensitive. Both Baidu Apollo and Autoware use type-aware strategies: (1) different tracker configurations per object type, and (2) penalized association when a detection’s type differs from an existing track, letting attackers disrupt tracking by manipulating perceived object types. Prior work~\cite{man2023person} examined camera misclassification but only at the detection layer.

In practice, an attacker can induce multiple propagation paths (\tblref{tbl:new_attacks}). For example, adversarial patches on a screen, or other attacks meeting the \texttt{Continuity} requirement, can trigger misclassifications at chosen times, causing track removal or ID switching. Compared with tracking hijacking~\cite{jia2020fooling,muller2022physical}, it generally requires lower \texttt{Precision} and \texttt{Continuity} than full object removal.

\begin{attack}[label=attack:latency_to_perception]{\newattackref{detection_latency_discrepancy}}
Multi-modal perception attack by latencies. Injecting processing delays in camera or LiDAR streams introduces temporal misalignment in sensor fusion, amplifying cross-modal discrepancies and degrading perception and downstream task performance.
\end{attack}

The other example connects the latency attacks with multi-modal sensors. Object detection and sensor fusion are highly sensitive to temporal misalignment, and such discrepancies can degrade perception and propagate downstream~\cite{muller2025investigating,liu2023slowlidar}.

Assuming fusion-based perception (e.g., LiDAR and camera), an adversary can inject latency into one modality to desynchronize it from the other and, knowing the pipeline, tune the delay for effect. Unlike spoofing or removal, latency attacks require far lower \texttt{Precision}, since no specific bounding box must be manipulated.


\myparagraph{Error amplification through downstream components} Small upstream perturbations can grow as they traverse modules such as tracking or motion prediction, effectively widening the attacker’s impact radius, as stated in \findingref{finding:prediction_as_amplifier}.

\begin{finding}[label=finding:prediction_as_amplifier]
    Object tracking and motion prediction components can effectively amplify object detection errors, which could overcome the limitations of certain physical sensor attacks in manipulating close-range objects. However, such attacks require high precision and continuity of the object detection attacks.
\end{finding}

This insight enables a broader class of attacks combining detection attacks with tracking and prediction manipulation (\newattackref{joint_detection_tracking_prediction}, \newattackref{acoustic_camera_tracking}), related to the physical object trajectory attack~\cite{lou2024first}, which targets detection and prediction simultaneously, and to our case study in \secref{sec:model}. The wider applicability and system-level impact of such combined attacks remain underexplored.


This category targets the propagation path from object detection errors to tracking and prediction failures. It places high demands on detection-stage errors, requiring strong \texttt{Precision} and \texttt{Continuity} to manipulate downstream modules. If feasible, it can trigger safety hazards more flexibly; e.g., inducing hard braking without deploying physical objects, unlike prior attacks~\cite{zhang2022adversarial,cao2022advdo}, or injecting fake objects at greater distances to remain stealthy. Such sophisticated strategies remain an important future direction.

\myparagraph{Hidden data dependencies across modules} Attackers can leverage certain data dependencies in the system architecture; for instance, many components depend on localization results (\findingref{finding:localization_and_beyond}).

\begin{finding}[label=finding:localization_and_beyond]
    Localization errors can propagate beyond the localization module, affecting components such as object detection, trajectory prediction, and planning, due to their shared dependence on accurate ego pose estimation.
\end{finding}

The GNSS spoofing LiDAR attack~\cite{li2021fooling} (\attackref{gnss_spoofing_lidar_correction}) is relevant here, leveraging GPS errors to corrupt LiDAR point clouds, though its feasibility remains questionable (\findingref{finding:precision_challenge}). Recognizing the broad dependency on localization outputs, we propose Attack \newattackref{prediction_by_localization}.

\begin{attack}[label=attack:localization_to_prediction]{\newattackref{prediction_by_localization}}
Prediction attack by localization error. By exploiting GPS spoofing or IMU-based attacks, the adversary introduces ego pose estimation errors. These localization inaccuracies distort the transformation of perception data into the global frame, leading to perturbed object trajectories and cascading prediction errors.
\end{attack}

Trajectory prediction attacks~\cite{zhang2022adversarial} show prediction modules can be highly sensitive to minor perturbations. An attacker could inject comparable perturbations (e.g., within 1 meter) through localization attacks instead. This still requires some \texttt{Precision} and \texttt{Continuity} in localization spoofing, but removes the physical objects used in prior work, making such attacks feasible as a fully remote threat.

\section{Proof-of-concept Experiments} \label{sec:experiments}

We conducted proof-of-concept experiments for Attacks \newattacksref{tracking_by_misclassification,prediction_by_localization} on the real-world nuScenes dataset~\cite{caesar2020nuscenes}, a representative driving scenario. Our pipeline combined BEVFusion~\cite{liang2022bevfusion} for multi-modal object detection, AB3DMOT~\cite{weng2020ab3dmot} for tracking, and Trajectron++~\cite{salzmann2020trajectron++} for trajectory prediction. These demonstrations showcase the conceptual potential of these attacks, though full feasibility assessment would require further real-world evaluation.

\begin{figure}
    \centering
    \includegraphics[width=0.98\linewidth]{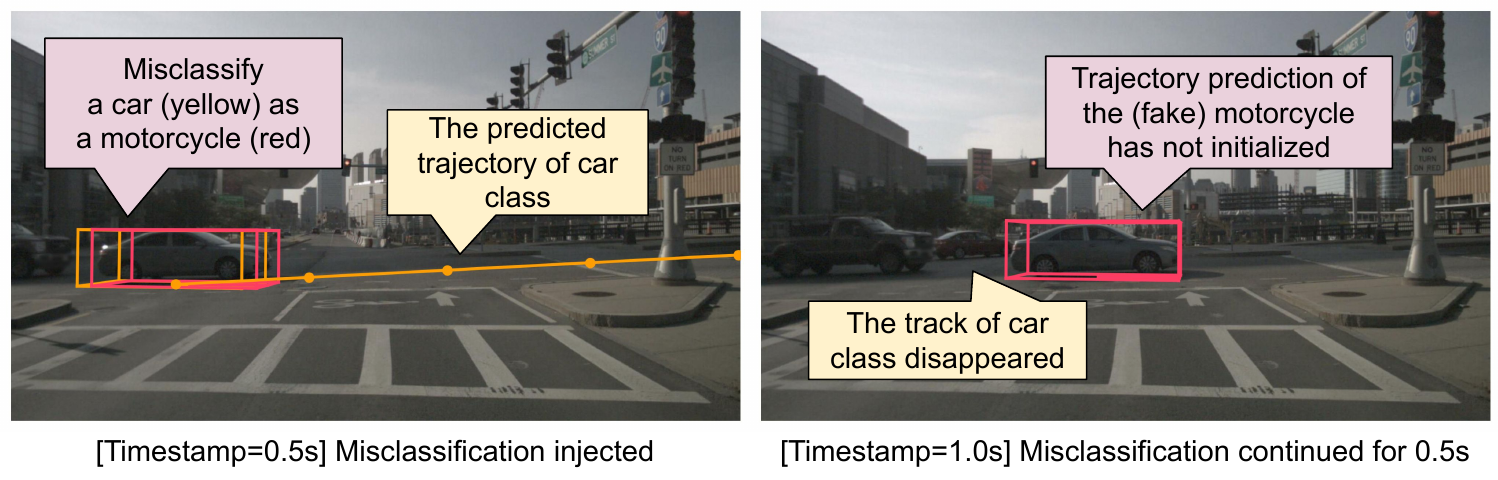}
    \caption{Attack \newattackref{tracking_by_misclassification}: Tracking attack by misclassification}
    \label{fig:misclassification_to_tracking}
\end{figure}

\myparagraph{Attack \newattackref{tracking_by_misclassification}: Tracking attack by misclassification}
\figref{fig:misclassification_to_tracking} shows the attack. Flipping a nearby object between the ``car'' (yellow) and ``motorcycle'' (red) labels makes AB3DMOT treat it as two distinct classes and initialize a new track for the misclassified motorcycle. Sustaining this for 0.5 s makes the tracker discard the original car track, erasing its trajectory history; this step requires the attack attribute \texttt{Continuity}, demonstrated in prior work~\cite{man2023person}. Since the new track has $< 0.5$ s of motion data, its prediction is not initialized, removing the true car’s prediction. Brief, low-precision perception errors thus cascade into tracking loss and prediction failure, though success depends on tracker/predictor implementations.

\begin{figure}
    \centering
    \includegraphics[width=0.98\linewidth]{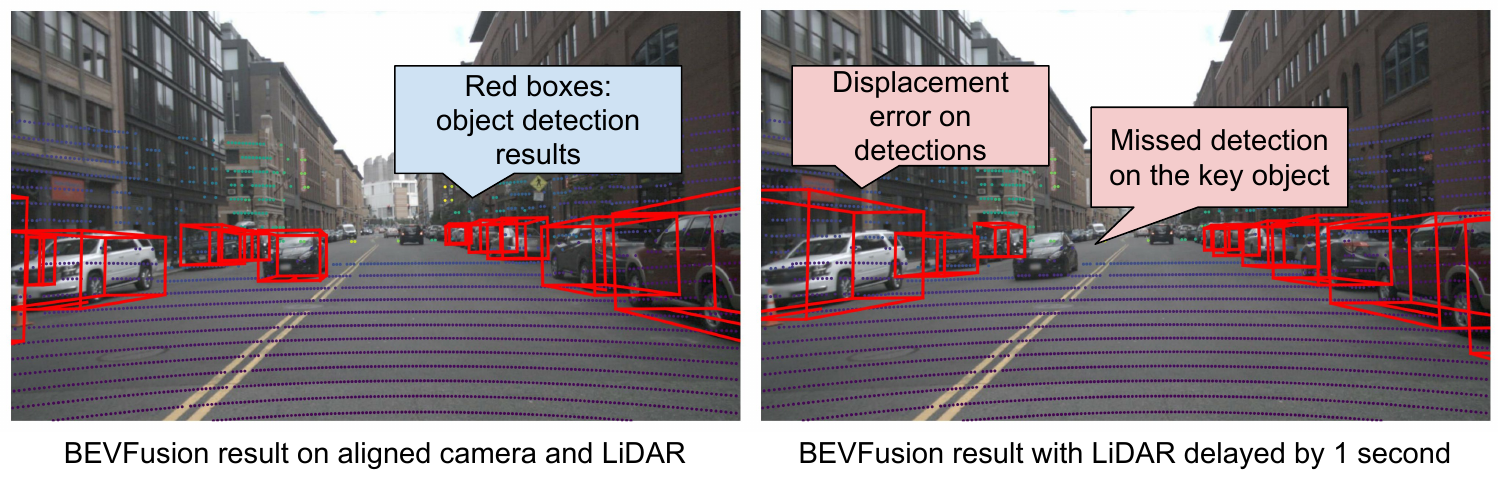}
    \caption{Attack \newattackref{detection_latency_discrepancy}: Multi-modal perception attack by latencies}
    \label{fig:latency_to_perception}
\end{figure}

\myparagraph{Attack \newattackref{detection_latency_discrepancy}: Multi-modal perception attack by latencies}
\figref{fig:latency_to_perception} shows the impact of object detection latency on BEVFusion, a widely used multi-modal detector. A 1-second delay in the LiDAR stream causes displacement and false negatives in fused results, including a missed vehicle just 10 m ahead. Prior work evaluated the negative impact of sensor misalignment~\cite{yu2023benchmarking}, typically assuming $<$ 0.5 s frame asynchrony. Adversarial attacks can induce larger delays, e.g., over 8 s in some cases~\cite{liu2023slowlidar,muller2025investigating}, leading to severe perception failures. The impact also varies by AI model; BEVFusion, for instance, is particularly sensitive to LiDAR latency~\cite{yu2023benchmarking}.


\begin{figure}[t]
    \centering
    \includegraphics[width=0.9\linewidth]{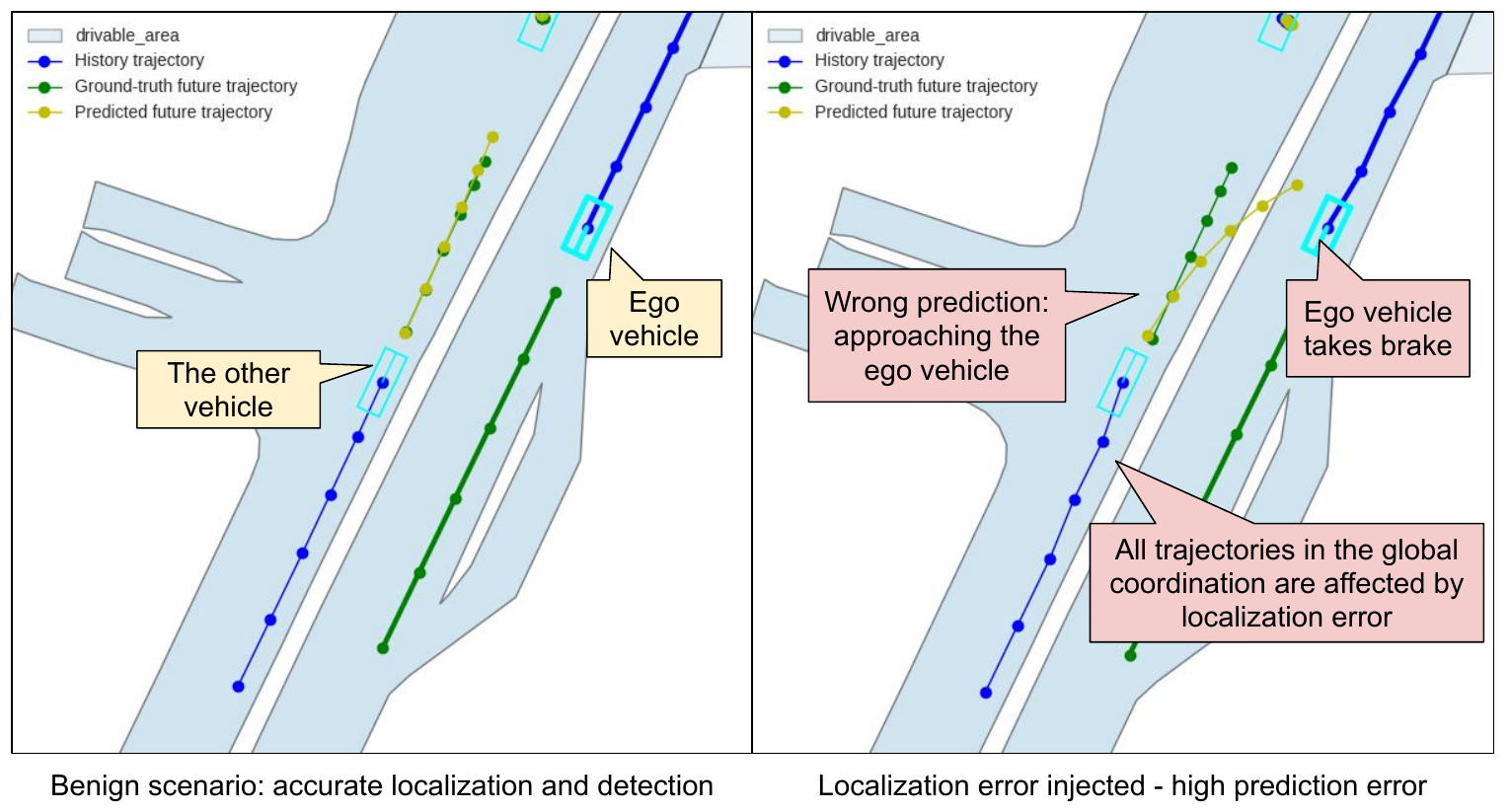}
    \caption{Attack \newattackref{prediction_by_localization}: Prediction attack by localization error}
    \label{fig:localization_to_prediction}
\end{figure}

\myparagraph{Attack \newattackref{prediction_by_localization}: Prediction attack by localization error}
We simulate the effect of ego localization error on downstream prediction by injecting perturbations ($<$ 1 meter) into the ego pose before detection and tracking. 
\figref{fig:localization_to_prediction} illustrates a representative case: the error is global, so all observed trajectories seem displaced in the ego vehicle's view. Under a certain perturbation, the prediction erroneously suggests the other vehicle will accelerate toward the ego vehicle, potentially triggering inappropriate planning responses. Subtle localization faults can thus mislead prediction even when perception and tracking themselves operate correctly.
\section{Pilot User Study} \label{sec:user_study}

To evaluate the usefulness of manually crafted \modelAbbr, we conducted an IRB-approved pilot study with 15 autonomous-vehicle security experts. Each reviewer evaluated a \modelAbbr for a paper they authored or knew well, via a questionnaire on correctness, expressivity, and usability. The study covers 13 of the 38 attack categories in \tblref{tbl:attack_summary}, and 87\% of reviewers were \emph{confident} or \emph{very confident} in their evaluation. Questionnaire design is in \appref{sec:questionnaire_design}.

\myparagraph{Correctness}
Reviewers found \modelAbbr to faithfully represent surveyed attacks. Across six correctness dimensions, including attack modeling, propagation paths, transition conditions, and assumptions, at least 90\% rated it \emph{Correct} or \emph{Largely correct}. Reported issues were limited to minor wording clarifications, which we fixed.

\myparagraph{Expressivity}
Reviewers considered \modelAbbr sufficiently expressive for qualitative system-level reasoning. All found the propagation-path representation sufficient, and 93\% approved the modeling of assumptions and constraints. Uncertainty modeling received the lowest rating (53\%), consistent with our choice to keep the model compact and qualitative rather than capture low-level execution details; \modelAbbr should thus complement, not replace, realistic experiments.

\myparagraph{Finding Quality}
The findings derived from \modelAbbr were rated highly: 93\% marked them \emph{Good} or \emph{Very good} in correctness, 87\% in relevance, and 80\% in clarity, suggesting our method produces well-grounded conclusions.

\myparagraph{Effort and Adoption}
Constructing a \modelAbbr appears practical: 73\% estimated an expert could create one for a familiar paper in under an hour, even without tooling. Adoption interest was strong: all reviewers would read survey papers organized with \modelAbbr, 87\% would adopt similar methods in their own work, and 67\% would contribute to a shared \modelAbbr library.
\section{LLM Generation of \modelAbbr}
\label{sec:llm}

Ultimately, we evaluated LLM-assisted \modelAbbr generation using GPT-4 (o3), Grok 3, and Gemini 2.5 Pro. For each attack, we provided the same prompt defining \modelAbbr nodes, attributes, transitions, conditions, and analysis objectives, plus the full text of the corresponding paper; the prompt template and model outputs are in \appref{sec:llm_analysis} for reproducibility. LLMs substantially automated literature extraction and graph construction but occasionally hallucinated nodes or unsupported propagation links. Across the evaluated papers, all generated graphs contained complete attack--error--impact chains; 65--70\% were error-free, 25--30\% contained only one or two errors, and no more than 5\% contained several. Expert, human-in-the-loop evaluation of node definitions, transition logic, feasibility assessments, and suggested attacks identified useful finer-grained propagation details and validated 80--100\% of the proposed stronger attacks. Overall, LLMs can produce useful first-draft \modelAbbrs and automate much of system-level security analysis, while expert verification remains necessary for semantic correctness and feasibility.


\section{Conclusion}

This study provides a systematic perspective on how sensor attacks propagate through autonomous vehicle pipelines, bridging isolated module analyses and real-world feasibility. We identify 8 key findings, propose 12 new attack opportunities enabled by unique error-propagation paths, and validate selected cases through proof-of-concept experiments on real data. A user study further shows \modelAbbr is effective for qualitative system-level reasoning, while LLM-assisted generation points to its future automation potential.

\bibliographystyle{IEEEtran}
\bibliography{references}

\appendices
\section*{Open Science}
\label{sec:open_science}

To support reproducibility and further research, we provide an anonymous artifact repository containing the materials used for SEPG construction and evaluation. The artifact includes materials for reproducing key results in the paper: manually generated SEPG graphs (saved as JSON metadata), LLM prompt templates used for SEPG generation, and the source code of proof-of-concept attacks. These materials are provided through the anonymous link \textbf{\url{https://anonymous.4open.science/r/VehicleSecSurvey-32A5}} during review and will be made publicly available in a permanent repository upon publication. The released artifacts are intended to facilitate transparent inspection of our graph-construction process, enable comparison between human and LLM-assisted SEPG generation, and support future extensions of SEPG-based analysis. The repository does not include exploit code or implementation details that would lower the barrier to reproducing physical sensor attacks.

\section*{Ethics Considerations}
\label{sec:ethics}

This work is a systematization of knowledge (SoK) study that surveys and analyzes sensor attacks on autonomous systems, including self-driving cars, ground robots, and UAVs. As such, it does not introduce novel low-level exploitation techniques, nor does it release tools or artifacts that could be directly misused to attack deployed systems. Instead, our contributions focus on organizing existing work, analyzing feasibility constraints, and highlighting overlooked error propagation paths for the purpose of advancing defensive research.

\myparagraph{No harmful experiments}
All proof-of-concept demonstrations in this paper were conducted using public datasets (e.g., nuScenes) or simulation environments. We did not perform experiments on live autonomous platforms, nor did we interfere with any operational systems, thereby ensuring no physical harm to people, vehicles, or environments.

\myparagraph{Human-subject study safeguards}
The user study involved voluntary participation by domain experts who reviewed SEPGs and provided feedback on their correctness, expressivity, and usability. The study was designed as a minimal-risk survey-based evaluation. We did not use deception, and survey responses are reported only in aggregate form without participant-identifying information.
The study was reviewed and approved by our institution’s Institutional Review Board (IRB) under exempt review for minimal-risk human-subject research involving survey procedures and benign behavioral interactions.

\myparagraph{Dual-use awareness}
We recognize that systematizing attack knowledge can inform both defenders and adversaries. To mitigate dual-use risks, we frame all analyses in terms of feasibility constraints and limitations, making clear where attacks are impractical or unrealistic. Our descriptions are kept at a conceptual level, without providing step-by-step instructions, hardware configurations, or exploitable code that could lower the barrier to malicious replication.

\myparagraph{Responsible disclosure}
Although our work does not identify new vulnerabilities in specific deployed systems, we recognize that some insights may generalize to existing platforms. Should any future findings point to undisclosed weaknesses in widely deployed systems, we are committed to following responsible disclosure practices in coordination with affected vendors.

\section{Additional Details of \modelAbbrs}

\begin{table*}[t]
\centering
\footnotesize
\caption{Attack/Error/Impact Nodes Modeled in SEPG}
\label{tbl:collection_nodes}
\begin{tabularx}{\textwidth}{lX}
\toprule
\textbf{Category} & \textbf{Nodes} \\
\midrule

\textbf{Attack Nodes} &
\textbf{Camera:} Physical patch, Patch by projector, Laser, Infrared light, Strong light;
\textbf{LiDAR:} LiDAR spoofing;
\textbf{Radar:} mmWave radar spoofing; 
\textbf{GNSS:} GNSS spoofing; 
\textbf{IMU:} Acoustic signal, Electromagnetic signal; 
\textbf{Physical objects:} Physical adversarial object, Physical natural object \\

\midrule

\textbf{Error Nodes} &
\textbf{Camera:} Camera detection false positives, Camera detection false negatives, Camera detection misclassification, Camera detection displacement, Camera image distortion, Camera misdetection (road sign), Camera misdetection (traffic signal), Camera misdetection (lane), Camera depth estimation error, Camera motion estimation error, Camera detection latency; \\
& \textbf{LiDAR:} LiDAR detection false positives, LiDAR detection false negatives, LiDAR detection latency, LiDAR image distortion, LiDAR motion estimation error; \\
& \textbf{Radar:} Radar detection false positives, Radar detection false negatives, Radar range measurement error, Radar angle measurement error, Radar speed measurement error; \\
& \textbf{GNSS:} GNSS deviation error; 
\textbf{IMU:} IMU reading error; \\
& \textbf{Perception:} Object detection false positives, Object detection false negatives, Object detection misclassification, Object detection displacement, Object detection latency, Depth estimation error, Tracking displacement, Tracking hijacking, Tracking broken tracks, Tracking lost track, Tracking ID switch, Trajectory prediction spoofed trajectory, Trajectory prediction displacement error, Trajectory prediction error, Coordination transformation error; \\
& \textbf{Localization:} Localization deviation error, Mapping error, Motion estimation error; \\
& \textbf{Planning:} Routing error, Planning error \\

\midrule

\textbf{Impact Nodes} &
Collision, Braking, Hard braking, Evasion maneuver, Off-road driving, Rule violation, Unexpected driving decisions, Drifting, Stalling, Lane change, Acceleration, Blocking other vehicles, Persistent intersection blocking \\

\bottomrule
\end{tabularx}
\end{table*}
\begin{table*}[t]
    \fontsize{6.5}{7.5}\selectfont
    \centering
    \renewcommand{\arraystretch}{1.2}
    \setlength{\tabcolsep}{2pt}
    
    \caption{Rules of constructing \modelAbbr.}
    \label{tbl:collection_rules}
    
    \begin{tabularx}{\textwidth}{|p{5cm}|X|}
    \noalign{\global\arrayrulewidth1pt}\hline\noalign{\global\arrayrulewidth0.4pt}
    \textbf{Rule} & \textbf{Description} \\
    \noalign{\global\arrayrulewidth1pt}\hline\noalign{\global\arrayrulewidth0.4pt}
    
    \textbf{R1: Attack nodes use the earliest attacker-controlled entry point} &
    Instantiate an attack node at the earliest attacker-controlled perturbation, rather than at a downstream system error. For example, use attack nodes such as physical patch, laser, LiDAR spoofing, or GNSS spoofing, instead of camera false positive or planning error. \\
    \hline
    
    \textbf{R2: Error nodes represent semantically wrong internal states} &
    Instantiate an error node whenever an internal CPS representation becomes semantically wrong, such as camera detection false negative, tracking hijacking, localization deviation error, or planning error. Such corruption should not be hidden inside transitions. \\
    \hline
    
    \textbf{R3: Module-boundary crossing creates a new error node} &
    Create a new error node whenever the perturbation crosses a subsystem boundary with different semantics, e.g., sensing $\rightarrow$ single-sensor perception, perception $\rightarrow$ tracking, tracking $\rightarrow$ prediction, motion estimation $\rightarrow$ localization, or localization/perception/prediction $\rightarrow$ planning. \\
    \hline
    
    \textbf{R4: Impact nodes are only physical or mission-level outcomes} &
    Use impact nodes only for physical-world or task-level consequences, such as collision, braking, off-road driving, rule violation, or blocking other vehicles. Internal software states such as planning error remain error nodes rather than impact nodes. \\
    \hline
    
    \textbf{R5: Default inheritance} &
    By default, the downstream error node inherits the formal attributes from upstream error nodes, unless a later rule updates them. The inherited attributes are limited to Precision, Continuity, Intensity, DynamicTarget, DynamicImpact, and DistanceImpactToTarget. \\
    \hline
    
    \textbf{R6: Use the finest node semantics supported by evidence} &
    When a paper provides specific semantics, preserve them in the node, e.g., traffic signal misdetection rather than a generic camera detection error. Use coarser node names only when the paper does not support finer-grained modeling. \\
    \hline
    
    \textbf{R7: Only six characteristics are formal attributes} &
    The only formal node attributes modeled in \modelAbbr\ are Precision, Continuity, Intensity, DynamicTarget, DynamicImpact, and DistanceImpactToTarget. Other information, such as camera-only dependence, fusion assumptions, map dependence, attacker knowledge, or scenario dependence, should not be encoded as attributes. \\
    \hline
    
    \textbf{R8: Other assumptions are string annotations} &
    Any property outside the six formal attributes should be attached as a string annotation to a node or transition. Examples include camera-only, LiDAR dominates fusion, no HD map, traffic-light scenario, black-box attacker, line-of-sight required, and only if recovery fails. These annotations notify the reader about assumptions not directly represented in the graph. \\
    \hline
    
    \textbf{R9: Uncertainty degrades precision} &
    If the attack has uncontrollable real-world factors that harm accuracy, cap the node's attribute \texttt{Precision} to \texttt{Medium}. If the paper only demonstrates generic disturbance or instability, further cap \texttt{Precision} to \texttt{Low}. \\
    \hline
    
    \textbf{R10: Precision requires demonstrated semantic controllability} &
    Assign \texttt{Precision}=\texttt{High} only if the paper demonstrates control over a specific semantic outcome, such as a chosen false class, chosen displacement, or chosen trajectory. Assign \texttt{Medium} if the attacker reliably influences one semantic dimension but cannot fully specify the outcome. \\
    \hline
    
    \textbf{R11: Per-frame success indicates continuity} &
    If the attack has a success rate under 90\% per frame, cap the node's attribute \texttt{Continuity} to \texttt{Medium}. If the paper only demonstrates one-shot or sporadic success, set \texttt{Continuity} to \texttt{Low}. \\
    \hline
    
    \textbf{R12: Stateful downstream modules can upgrade continuity} &
    A downstream stateful module can upgrade \texttt{Continuity} if it preserves the error over time. For example, object tracking can upgrade \texttt{Continuity} from \texttt{Medium} to \texttt{High}, while localization memory or planner memory can preserve transient upstream errors. \\
    \hline
    
    \textbf{R13: Tracking upgrades continuity} &
    Object tracking by definition can upgrade node attribute \texttt{Continuity} from \texttt{Medium} to \texttt{High}. \\
    \hline

    \textbf{R14: Prediction extends positional error} &
    Prediction extrapolates location errors into a longer-term future, potentially shrinking \texttt{DistanceImpactToTarget} from \texttt{Medium} to \texttt{Close}. More generally, downstream decision-layer nodes may reduce \texttt{DistanceImpactToTarget} when they move the corruption closer to physical actuation. \\
    \hline
    
    \textbf{R15: Tracking/prediction attacks require continuity and dynamic upstream errors} &
    The transition to tracking error requires \texttt{Continuity} $>$ \texttt{Medium} while transition to prediction error requires \texttt{Continuity} $>$ \texttt{High}. Both transitions require \texttt{DynamicImpact}. \\
    \hline
    
    \textbf{R16: Sensor fusion is modeled as a transition, not as an attribute} &
    Model fusion as a special transition that merges multiple sensor-error nodes into one detection/localization error node. Assumptions such as camera dominates fusion, LiDAR dominates fusion, weak cross-checking, or fallback mode should be written as string annotations on that transition rather than encoded as node attributes. \\
    \hline

    \textbf{R17: Transitions follow semantic consumption} &
    Add an edge only if the downstream module actually consumes the corrupted upstream semantics. For example, object detection displacement can propagate to tracking hijacking if the tracker uses the displaced detection, while camera false negative usually does not directly propagate to routing error. \\
    \hline
    
    \textbf{R18: Localization transitions require pose-relevant upstream errors} &
    A transition to localization deviation error is allowed only if the upstream node affects ego pose estimation, relative motion estimation, inertial integration, scan matching, map alignment, or other pose-relevant signals. Examples include GNSS deviation error, IMU reading error, camera motion estimation error, and LiDAR motion estimation error. \\
    \hline
    
    \textbf{R19: Planning transitions require decision relevance} &
    A transition to planning error is allowed only if the upstream node can alter a planner-consumed semantic variable, such as obstacle existence, obstacle position, object class, lane geometry, traffic signal/sign state, future trajectory, ego pose, or route choice. \\
    \hline

    \textbf{R20: Scenario dependency in transitions} &
    Transitions to planning error conditioned on object geometry or ego motion/intent are labeled as \emph{scenario-dependent}, i.e., feasible only in certain scenes. This label should be attached as a string annotation to the transition, for example traffic-light scenario, lane-change scenario, close-range interaction, or no HD map. \\
    \hline

    \textbf{R21: Conjunctive transitions are explicit} &
    If a downstream error requires multiple upstream conditions simultaneously, model the propagation as an AND-style transition rather than separate independent edges. For example, a fused false negative may require both camera false negative and LiDAR false negative; collision may require both localization deviation error and missing obstacle avoidance opportunity. \\
    \hline
    
    \textbf{R22: Recovery mechanisms suppress or constrain transitions} &
    If downstream modules are likely to correct the upstream error, either omit the transition or annotate it with a string such as \emph{only if recovery fails}. Typical recovery mechanisms include cross-sensor consistency checks, map matching, temporal filtering, and safety fallback logic. \\
    \hline
    
    \textbf{R23: Close-distance requirement for collision/brake} &
    The transition to physical impact nodes like collision or braking requires \texttt{DistanceImpactToTarget} $=$ \texttt{Close} and \texttt{DynamicTarget}. \\
    \hline
    
    \textbf{R24: Planning-to-impact transitions require actuation relevance} &
    A transition from planning error to a physical impact node is allowed only if the planning error can directly alter commanded motion and is not effectively neutralized by downstream control or safety overrides. Additional assumptions, such as failed avoidance or close-range interaction, should be attached as string annotations to the transition. \\
    \hline
    
    \textbf{R25: Attacker and system assumptions are annotations, not graph structure} &
    Properties such as black-box attacker, white-box attacker, line-of-sight required, camera-only, LiDAR-only, no HD map, or precise timing required should be recorded as string annotations attached to nodes or transitions, rather than as extra nodes or formal attributes. \\
    \hline
    
    \textbf{R26: Evidence strength is an annotation} &
    Evidence strength, such as demonstrated in full system, demonstrated in subsystem, simulation only, or mechanistic inference, should be recorded as a string annotation on nodes or transitions, rather than as a formal attribute. \\
    \hline

    \textbf{R27: Closed-loop effects are annotations} &
    Whether downstream system behavior damps, preserves, or amplifies the upstream error should be recorded as a string annotation on the corresponding transition or node. This information is important for interpretation, but it is not one of the formal node attributes. \\
    
    \noalign{\global\arrayrulewidth1pt}\hline\noalign{\global\arrayrulewidth0.4pt}
    \end{tabularx}
\end{table*}

\begin{figure*}[h]
    \centering
    \includegraphics[width=0.99\linewidth]{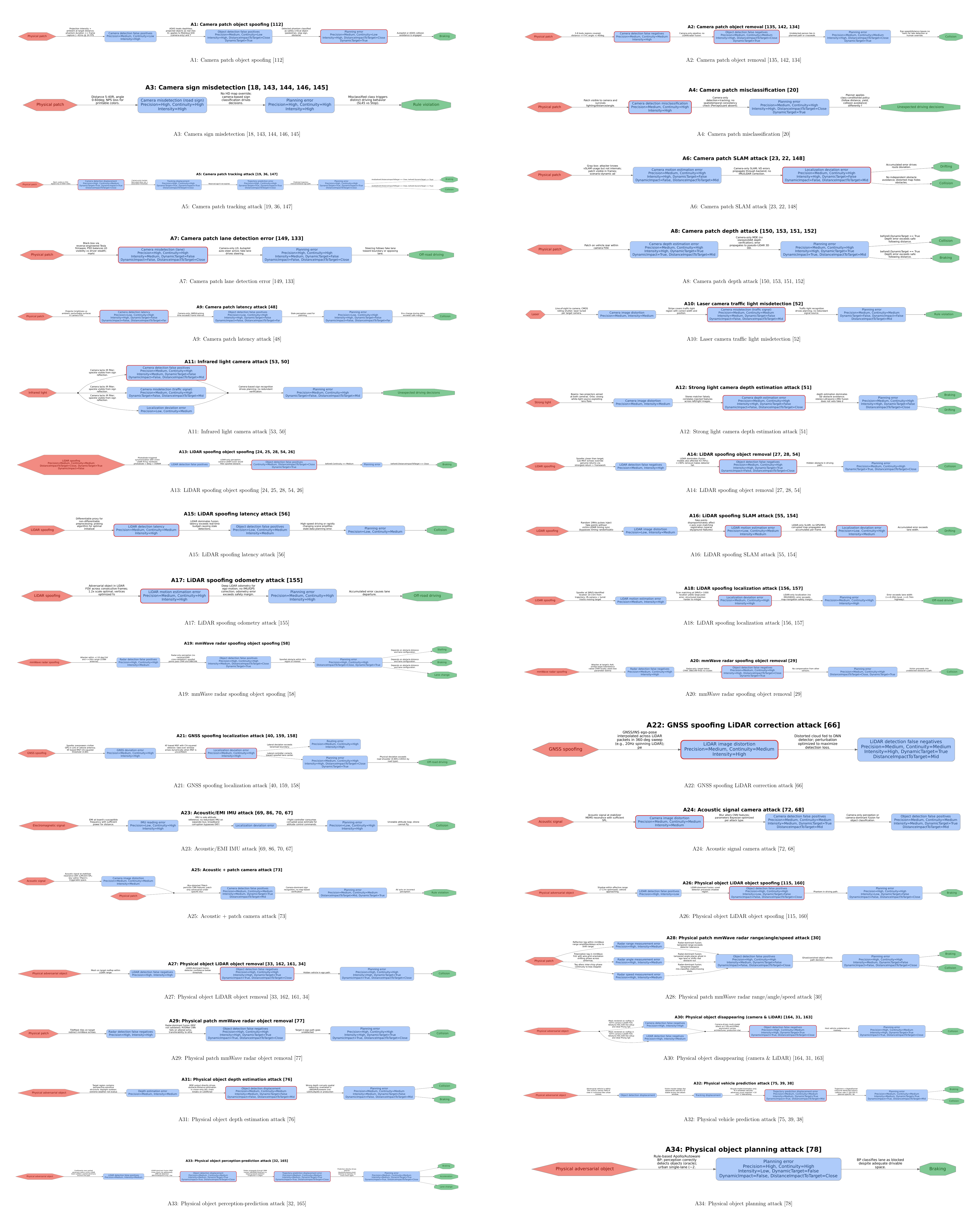}
    \caption{Visualized \modelAbbrs of attacks in \tblref{tbl:attack_summary}}
    \label{fig:all_sepg_graphs}
\end{figure*}

This section provides additional details on the construction and representation of \modelAbbrs. \tblref{tbl:collection_nodes} summarizes the nodes collected from the surveyed literature, while \tblref{tbl:collection_rules} presents the propagation rules used to connect these nodes and capture how attack-induced errors propagate across system components toward downstream impacts. Together, these tables provide the underlying elements used to instantiate the \modelAbbrs for the attacks summarized in \tblref{tbl:attack_summary}. For completeness, \figref{fig:all_sepg_graphs} visualizes the expanded \modelAbbrs for these attacks, exposing the intermediate states and propagation paths that may be omitted from the more compact representations in the main text.

\section{Using LLMs to generate \modelAbbrs}
\label{sec:llm_analysis}

We evaluate the feasibility of using LLMs to automate the generation and analysis of \modelAbbrs. Specifically, we test GPT-4 (o3), Grok 3, and Gemini 2.5 Pro on the same attack cases using identical prompts. We select these models because they support advanced reasoning and sufficiently large context windows to process full research papers. We note that LLM-generated \modelAbbrs necessarily involve abstraction and interpretive reasoning and may therefore vary across models and prompt formulations. Our goal is to evaluate whether contemporary LLMs can provide useful and reproducible assistance for constructing \modelAbbrs, rather than to establish any particular model as an authoritative analyzer.

We begin each query with the following role-playing system prompt:

\begin{lstlisting}[language={},frame=single,frameround=ffff,
showstringspaces=false,xleftmargin=5pt,xrightmargin=5pt]
You are an expert in cyber-physical system security. 
You are responsible for understanding and analyzing 
the results from academic papers.
\end{lstlisting}

The prompt then provides the definition of \modelAbbr, largely following \secref{sec:model}. In particular, it explains the supported node types, node attributes, transitions, and the use of preconditions and postconditions for describing how an error can propagate from one system component to another. We additionally provide illustrative examples of individual nodes and transitions to clarify the expected representation format. However, we deliberately do not provide complete human-generated \modelAbbr examples as few-shot demonstrations. This prevents the LLM from simply imitating the experts' graphs and allows us to independently assess its ability to derive the propagation structure from the source paper.

After these instructions, we provide the full text of the research paper being analyzed. Depending on the LLM service, the paper is supplied either by uploading its PDF or by providing a URL that the service can directly retrieve and parse. The same paper content and prompt structure are used across the three models. The main analysis request is:

\begin{lstlisting}[language={},frame=single,frameround=ffff,
showstringspaces=false,xleftmargin=5pt,xrightmargin=5pt]
**Major goal**: Next, please read the paper in the link 
given below, and generate the corresponding System Error 
Propagation Graph (\modelAbbr). Please be detailed in the 
definition of nodes and transitions (along with conditions 
over attributes). Think deeply and critically.

You can also share your findings of:
* Whether the attack is feasible and easy to reproduce.
* If the attack does not appear to be immediately feasible, 
  which specific transition steps within the system 
  architecture constrain or challenge the feasibility.
* Insights in designing stronger attacks by removing the 
  feasibility constraints, or designing effective mitigation 
  methods by leveraging the challenges of attack feasibility.
\end{lstlisting}

Thus, each LLM is asked to produce four main classes of outputs: (1) the \modelAbbr node set and associated attributes, (2) the transition set together with relevant preconditions and postconditions, (3) an assessment of attack feasibility and the propagation steps that constrain it, and (4) possible stronger attack or mitigation strategies suggested by the identified constraints. The generated outputs are retained in their original Markdown form rather than being manually normalized before comparison.

To systematically evaluate the outputs, we construct a database\footnote{The database link has been omitted to comply with the double-blind review process.} containing the LLM-generated analyses for the papers listed in Table~\ref{tbl:attack_summary}, together with the prompt templates used to generate them. The database and prompts will be released as open source upon publication.

For a human-in-the-loop comparison across models, we use the attack by Zhang et al.~\cite{zhang2022adversarial} as a representative case study. The three LLMs independently analyze the same paper under the prompt described above, and we compare their outputs along four dimensions: (1) node definitions, (2) transition sets, (3) attack feasibility assessments, and (4) proposed strategies for designing stronger attacks. This experiment is intended as a proof of concept for LLM-assisted \modelAbbr generation and to characterize the consistency and limitations of such automation across different models.

\newcolumntype{S}{>{\raggedright\arraybackslash}p{1.9cm}} 
\newcolumntype{M}{>{\raggedright\arraybackslash}p{4.2cm}} 
\newcolumntype{T}{>{\raggedright\arraybackslash}p{2.8cm}} 
\newcolumntype{Y}{>{\raggedright\arraybackslash}X}        

\begin{table*}[t]
\centering
\scriptsize
\setlength{\tabcolsep}{4pt}
\renewcommand{\arraystretch}{1.2}
\caption{Objective metrics for evaluating the effectiveness and efficiency of LLM-generated SEPGs.
Abbreviations: NN = None, FW = Few (1$\sim$2), SV = Several (2$\sim$4), MN = Many ($\geq$5), MB = Maybe, MG: Marginal, SG = Significant, PR = Poor, FA = Fair, GD = Good, EX = Excellent.}
\begin{tabularx}{\textwidth}{@{} S M T Y @{}}
\toprule
\textbf{Dimension} & \textbf{Metric} & \textbf{Type/Val} & \textbf{Why it matters} \\
\midrule

\multirow{2}{*}{\textbf{\makecell[l]{\textbf{Structured}\\\textbf{Fidelity}} }}
  & \textbf{Completeness of LLM SEPG (COMP)}
  & \makecell[l]{Qual/\\Val=\{YES, NO\}}
  &  Ensures the extracted graph is complete relative to the paper and human version, avoiding missed components in node set $\mathcal{V}$, \texttt{Attributes}, and \texttt{Transitions}. \\
& \textbf{LLM SEPG Error Range (ER)} 
  & \makecell[l]{Quant/\\Val=\{NN, FW, SV, MN\}}
  & Captures graph correctness and schema discipline; explicitly counts structural mistakes (wrong/missing node/transition types, malformed hyper-edges).\\
\hline
\multirow{2}{*}{\textbf{Usability 
}}
  & \textbf{Complementary Perspectives (CP)}
  & \makecell[l]{Qual/\\Val=\{NN, MB, MG, SG\}}
  &  Whether the LLM SEPG provides insights or complementary perspectives that the user would have otherwise missed when analyzing the paper manually. \\
& \textbf{Human Error Correction (HEC)} 
  & \makecell[l]{Quant/\\Val=\{NN, FW, SV, MN\}}
  & Captures how helpful the LLM-generated output is in identifying or correcting mistakes in  manually generated SEPG.\\

\hline
\multirow{2}{*}{\textbf{\makecell[l]{\textbf{Expert-Judged}\\\textbf{Reasoning }} }}
  & \textbf{Feasibility Analysis Quality (FAQ)}
  & \makecell[l]{Qual/\\Val=\{PR, FA, GD, EX\}}
  &  An expert-rated stronger-attack hypothesis evaluation. \\
& \textbf{Stronger Attack Generation (SAG)} 
  & \makecell[l]{Quant/Fraction}
  & An expert-judged stronger-attack hypothesis evaluation. Reasonable attack vectors over the total number of attacks suggested.\\

\bottomrule
\end{tabularx}

\label{tab:llm_metrics}
\end{table*}
\newcolumntype{L}{>{\raggedright\arraybackslash}p{2.0cm}} 
\newcolumntype{S}{>{\centering\arraybackslash}p{1.1cm}}   
\newcolumntype{Q}{>{\centering\arraybackslash}p{0.9cm}}   
\newcolumntype{Y}{>{\raggedright\arraybackslash}X}        

\begin{table*}[t]
\tiny
\centering
\caption{Evaluation matrix for LLM-generated SEPG across papers. }
\label{tab:llm_analysis}
\renewcommand*{\arraystretch}{1.1}
\setlength{\tabcolsep}{2pt}
\begin{tabularx}{\textwidth}{|Q|S|S|Q|Q|Q|Q|Q|Y|}
\noalign{\global\arrayrulewidth1pt}\hline\noalign{\global\arrayrulewidth0.4pt}
\thead{Ref} 
& \thead{Sensor} 
& \thead{COMP} 
& \thead{ER} 
& \thead{CP} 
& \thead{HEC} 
& \thead{FAQ} 
& \thead{SAG} 
& \thead{Notes} \\
\noalign{\global\arrayrulewidth1pt}\hline\noalign{\global\arrayrulewidth0.4pt}


\cite{zhou2022doublestar} & \CameraIcon & YES & NN & SG & NN & EX & 6/6 &
Detailed attacks; feasibility analysis captures constraints comprehensively. \\
\hline

\cite{yan2022rolling} & \CameraIcon & YES & NN & SG & NN & EX & 5/5 &
Two natural extensions added to the attack nodes: laser-induced color-balance bias and autofocus disturbance. \\
\hline

\cite{wang2021can} & \CameraIcon & YES & NN & SG & NN & EX & 4/5 &
Well-documented attack details and transitions, with clear impact nodes. \\
\hline

\cite{song2023acero} & \CameraIcon & YES & NN & SG & NN & EX & 3/4 &
Not a pure attack paper (works as an adversarial driving behavior detector), but the SEPG demonstrated sound causal reasoning with rich details. \\
\hline

\cite{shamsi2024wip} & \CameraIcon  & YES & NN & MG & NN & EX & 3/4 &
Feasibility analysis is insightful and detailed. \\
\hline

\cite{sato2021dirty} & \CameraIcon & YES & FW & MB & NN & GD & 5/5 &
\texttt{ScenarioDependence}=Medium is questionable (under-lighting scenarios may break lane detection, so the attribute should be Low-Medium). Insightful paths to stronger attacks. \\
\hline

\cite{nassi2020phantom} & \CameraIcon & YES & NN & MG & NN & EX & 4/4 &
Insightful stronger attacks; graph notes higher success in low-lux and on high-albedo surfaces. \\
\hline

\cite{muller2025investigating} & \CameraIcon & YES & NN & MG & NN & PR & 6/7 &
Feasibility analysis missed key constraints: attacker’s access to victim camera feed, projector availability, and white-box OD knowledge. Clear attack path summaries; quantified intensity attributes. \\
\hline

\cite{chen2024adversary} & \CameraIcon & YES & FW & MG & NN & FA & 3/4 &
\texttt{DistanceImpactToTarget} attribute should be Low–Medium. End-to-end error decomposition into segments with detailed, correct attack paths. \\
\hline

\cite{muller2022physical} & \CameraIcon & YES & FW & MG & NN & EX & 3/3 &
\texttt{DistanceImpactToTarget} should be lower on prediction/planning nodes. Provided more detailed tracking errors and more impact nodes dependent on scene. \\
\hline

\cite{man2023person} & \CameraIcon & YES & FW & MB & NN & EX & 3/3 &
\texttt{DistanceImpactToTarget} should be lower on prediction/planning nodes. LLM proposed improvements on stealthiness (stronger-attack angle). \\
\hline

\cite{jing2021too} & \CameraIcon & YES & NN & MB & NN & EX & 3/3 &
LLM added more detailed lane-detection errors and impact articulation. \\
\hline

\cite{jia2020fooling} & \CameraIcon & YES & NN & MB & NN & GD & 3/3 &
\texttt{DistanceImpactToTarget} should be lower on prediction/planning nodes. \\
\hline

\cite{ji2021poltergeist} & \CameraIcon+\ImuIcon & YES & SV & MB & NN & FA & 3/3 &
Precision should be lower; impact is perception-level (not physical); did not mention image patch. \\
\hline

\cite{guo2024invisible} & \CameraIcon & YES & NN & MB & FW & GD & 4/4 &
\texttt{DynamicImpact} should be False. \\
\hline

\cite{eykholt2018robust} & \CameraIcon &  YES & NN & MB & NN & GD & 4/4 &
\texttt{DynamicImpact} should be False. LLM added an additional impact node. \\
\hline

\cite{cheng2022physical} & \CameraIcon & YES & NN & MB & NN & EX & 5/5 &
Paper not systematic, but LLM provided more detailed structure. \\
\hline

\cite{cao2022advdo} & \CameraIcon & YES & NN & MB & NN & EX & 4/4 &
More detailed errors and impact nodes (possibly too granular). Stronger attacks presented as a bottleneck analysis. \\
\hline

\cite{jeong2023rocking} & \ImuIcon & YES & NN & MB & NN & FA & 1/2 &
The paper does not show a systematic view of the IMU attack. \\
\hline

\cite{lou2024first} & \LidarIcon & YES & NN & SG & FW & GD & 2/2 &
Added \emph{E4 Agent not predicted} — complemented the manual graph with a plausible new error node. \\
\hline

\cite{shen2020drift} & \LidarIcon & YES & FW & MG & NN & GD & 0/1 &
Missing victim tracking node in the node set. \\
\hline

\cite{sun2020towards} & \LidarIcon & YES & FW & MB & NN & FA & 1/1 &
Spurious fusion node: \emph{E3 E-FusionDet\_Inc | residual inconsistency} should not be present. \\
\hline

\cite{zhu2021can} & \LidarIcon & YES & FW & MG & FW & GD & 1/1 &
A node ``Mismatch with camera'' was wrongly introduced. LLM proposed a ``position/size bias'' error node that makes sense. \\
\hline

\cite{hallyburton2022security}
& \CameraIcon /\LidarIcon
& YES & NN & SG & FW & FA & 2/3
& The LLM SEPG provides insights into SAG (planning errors can be intensified when the number of tracking objectives is limited). \\
\hline

\cite{cao2019adversarial}
& \LidarIcon
& YES & FW & MB & NN & GD & 3/3
& Missed AV-freezing impact node. Also added a bogus cam–LiDAR fusion; fusion is a defense---probably cross-paper bleed. \\
\hline

\cite{cao2021invisible} & \LidarIcon & YES & FW & MB & NN & GD & 2/2 &
Graph wrongly included tracking in nodes; ``hard brake'' impact cannot be realized under this threat model. \\
\hline

\cite{fukunaga2024random} & \LidarIcon & YES & FW & SG & NN & EX & 2/3 &
Attack model cannot yield hard braking (only z-axis localization affected). Nice catch: “drift not reset by loop closure” as an effectiveness condition. Suggests prior map building as a stronger attack, but it is too challenging \\
\hline

\cite{jin2023pla} & \LidarIcon & YES & FW & MB & NN & GD & 1/2 &
Cross-modal mismatch was not an attack in the paper but appeared in the graph. Some useful stronger-attack insights. \\
\hline

\cite{li2021fooling} & \LidarIcon & YES & FW & MB & NN & GD & 2/2 &
False node: \emph{E4 E-FusionDet\_Inc | mismatch with camera}, since no camera is introduced. \\
\hline

\cite{liu2023slowlidar} & \LidarIcon & YES & FW & MB & NN & EX & 1/1 &
``Cross-modal timing skew'' should not be an error node. Transition summary in the graph is cleaner than the paper's. \\
\hline

\cite{zhang2022adversarial} & \LidarIcon/\CameraIcon & YES & NN & SG & NN & EX & 5/5 &
SEPG shows fine-grained results and insightful feasibility analysis. \\
\hline

\cite{chen2023metawave} & \RadarIcon & YES & FW & MB & NN & FA & 3/3 &
Fusion disagreement node was wrongly added; planning should not be reflected (though planning impacts are a natural downstream effect). Insightful stronger attacks suggested. \\
\hline

\cite{hunt2024madradar} & \RadarIcon & YES & NN & MG & NN & FA & 1/2 &
Good capture of two key preconditions: eavesdropping SNR $>\tau$ and robustness to antenna rotation. \\
\hline

\cite{sun2021control} & \RadarIcon & YES & NN & MB & NN & FA & 1/2 &
Looks correct; similar attack vector to related work \cite{hunt2024madradar}. \\
\hline

\cite{zhu2023tilemask} & \RadarIcon & YES & FW & MB & NN & GD & 2/2 &
\texttt{DynamicImpact=True} should be \texttt{False} because the adversarial sign is fixed on the victim. \\
\hline

\noalign{\global\arrayrulewidth1pt}\hline\noalign{\global\arrayrulewidth0.4pt}
\end{tabularx}
\end{table*}

To further evaluate the quality of the LLM-generated \modelAbbrs, we examined the Markdown outputs in our constructed database for each paper and conducted a cross-paper analysis. Given the release of GPT-5 Thinking, we regenerated all \modelAbbrs with this model to leverage its stronger reasoning capability. Domain experts (the authors) adjudicated the results. We first introduce the evaluation metrics in Table~\ref{tab:llm_metrics}: structural fidelity (COMP, ER), usability (CP, HEC), and expert-judged reasoning (FAQ, SAG), measured qualitatively and quantitatively. Table~\ref{tab:llm_analysis} then applies these metrics across papers as a human evaluation matrix.

Based on Tables~\ref{tab:llm_metrics} and~\ref{tab:llm_analysis}, LLM-generated \modelAbbrs provide a complete and structured baseline for analysis. Completeness (COMP) is 100\%, meaning every paper yields an attack$\rightarrow$error$\rightarrow$impact chain with attributes and transitions. Structural defects are limited: approximately 65--70\% of cases exhibit NN (no structural errors), 25--30\% exhibit FW (few, 1--2), and $\leq$5\% exhibit SV (several, 2--4). This distribution indicates that reviewers begin from a consistent, largely correct graph rather than from scratch, which reduces extraction effort and enables parallel comparison across studies.

Beyond extraction, the LLM confers analytical value. Complementary Perspectives (CP) are frequently nontrivial, often MG/SG, and Stronger Attack Generation (SAG) demonstrates high precision where reported, with typical hit rates of ~80–100\% (for example, 6/6, 5/5, 4/5, 3/4) validated by experts. In practice, these outputs surface testable hypotheses, including scenario levers, necessary preconditions, and propagation bottlenecks, while keeping speculative suggestions clearly delineated from claims grounded in the source paper.

Expert-judged Feasibility Analysis Quality (FAQ) trends high, with GD/EX common, suggesting that the synthesized chains typically respect realistic constraints. The principal limitation is modest Human Error Correction (HEC), mostly NN with some FW, indicating that the LLM is not yet a dependable automatic checker of human mistakes. A suitable usage pattern is therefore to treat the LLM \modelAbbr as a faithful first draft that accelerates review and ideation, followed by a brief human quality assurance pass, for example, cite-to-claim gating and attribute guards, to lock down fidelity.

\noindent
\textbf{Discussion}. Overall, LLM-generated \modelAbbrs give a complete and mostly correct starting point for analysis, and they often surface useful hypotheses that experts can test. With a brief human review, the graphs help organize attack, error, and impact reasoning, highlight feasibility constraints, and suggest stronger attacks that remain tied to evidence in the paper.

These findings are provisional. Results can change with model updates and can vary across repeated runs. To keep results consistent over time, record the model name and date for each analysis, consider generating two or three runs and using simple consensus, and re-benchmark key papers after major model updates.

\section{Expert Questionnaire Design}
\label{sec:questionnaire_design}

We designed an expert questionnaire to evaluate SEPG along three dimensions: model correctness, model expressivity, and usability and construction effort. Each participant was asked to evaluate the SEPG constructed for a specific research paper. Before answering the questionnaire, participants were given a brief introduction to SEPG and asked to review the corresponding SEPG provided by us. The questionnaire contained the following questions.

\paragraph{Module 1: Model Correctness.}
Correctness evaluates whether the constructed graph faithfully represents the claims and demonstrated results of the original paper.

\begin{enumerate}
    \item \textbf{Rate the correctness of the SEPG drafted for the research paper.}

    Participants separately evaluated:
    \begin{itemize}
        \item Physical sensor attack modeling;
        \item Error node attributes;
        \item Attack impact modeling;
        \item Error propagation paths;
        \item Transition conditions; and
        \item Attack assumptions.
    \end{itemize}

    The rating scale ranged from \emph{Incorrect} to \emph{Correct}, with two intermediate levels representing substantial or minor issues.

    \item \textbf{Describe any specific inaccuracies or missing information observed.} (Optional)
\end{enumerate}

\paragraph{Module 2: Model Expressivity.}
Expressivity evaluates whether SEPG is sufficiently rich to represent important details of the attack.

\begin{enumerate}
    \setcounter{enumi}{2}

    \item \textbf{Rate the expressivity of SEPG in the following aspects:}
    \begin{itemize}
        \item The error propagation path from the sensor error to the attack impact;
        \item Constraints or challenges that must be overcome for a successful attack;
        \item Uncertainties affecting attack success; and
        \item Assumptions about the system or environment.
    \end{itemize}

    Participants rated each aspect from not represented to sufficient and generalizable.

    \item \textbf{Describe any limitations in expressivity that were observed.} (Optional)

    \item \textbf{Which current aspects of SEPG need the most improvement?} (Select all that apply)
    \begin{itemize}
        \item Quantitative evaluation of attack success;
        \item Detailed metrics evaluated on realistic systems;
        \item Modeling how defensive approaches address attacks;
        \item Clarity of the current node, edge, and attribute design;
        \item Clarity of end-to-end propagation logic; and
        \item Other.
    \end{itemize}

    \item \textbf{Which additional aspects should be added to future versions of SEPG?} (Select all that apply)
    \begin{itemize}
        \item Additional attack vectors, including power-supply attacks, software failures, and multi-agent attacks;
        \item More explicit modeling of defensive mechanisms;
        \item Greater support for quantitative or probabilistic reasoning;
        \item Greater support for system-specific or real-world deployment constraints; and
        \item Other.
    \end{itemize}
\end{enumerate}

\paragraph{Module 3: Usability and Construction Effort.}
Usability evaluates whether SEPG helps readers understand security papers and facilitates the discovery of new attack or defense ideas.

\begin{enumerate}
    \setcounter{enumi}{6}

    \item \textbf{As a domain expert, how much effort would it take to create an SEPG for the evaluated paper?}

    The choices were a few minutes, approximately one hour, several hours, or more than a few hours.

    \item \textbf{Would templates, tooling, or a library of modularized attack components significantly reduce the construction effort?}

    Participants evaluated the usefulness of:
    \begin{itemize}
        \item GUI tools for simplifying the specification of attributes and conditions;
        \item Reusable propagation rules for common situations, such as propagation from object detection to tracking and prediction;
        \item A reference library of pre-built and validated SEPGs; and
        \item LLM-based tools that process paper content to generate SEPGs through in-context learning or retrieval.
    \end{itemize}

    Each mechanism was rated from \emph{Not helpful} to \emph{Very helpful}.

    \item \textbf{After reviewing the findings derived from the SEPG, how would you rate their quality?}

    Participants separately evaluated:
    \begin{itemize}
        \item \textbf{Correctness}: whether the findings make valid claims regarding attack limitations or potential new attacks or defenses;
        \item \textbf{Clarity}: whether the findings are easy to understand; and
        \item \textbf{Relevance to the SEPG}: whether the findings are connected to the elements represented in the SEPG.
    \end{itemize}

    Each aspect was rated from \emph{Not good} to \emph{Very good}.

    \item \textbf{How confident are you in your evaluation?}

    Participants selected among five confidence levels ranging from no confidence to very confident.

    \item \textbf{What is your major concern if you are not confident about your evaluation?} (Optional)

    \item \textbf{Would you like to adopt the SEPG approach in any of the following ways?} (Select all that apply)
    \begin{itemize}
        \item Reading survey or SoK papers organized using a similar graph-based methodology;
        \item Adopting similar approaches for security analysis in education or research;
        \item Contributing to a shared library of SEPG models; and
        \item Other.
    \end{itemize}
\end{enumerate}

\end{document}